\documentclass[epj]{svjour}
\usepackage{amsmath}
\usepackage[final]{graphicx}
\usepackage{color}
\usepackage{bm}
\def\kap{\ensuremath{\boldsymbol\kappa}}
\newcommand{\vct}[1]{\mathbf{#1}}
\newcommand{\pr}{{\bf q}\cdot{\bf r}_s}
\newcommand{\prt}{{\bf q}(t)\cdot{\bf r}_s}
\newcommand{\ti}{e^{\Omega^\dagger t}}

\newcommand{\phis}{\Phi_{\vct{q}}^{s}(t)}

\newcommand{\phise}{\Phi_q^{s(e)}(t)}

\begin{document}
\title{Tagged-particle motion in glassy systems under shear: Comparison of mode coupling theory and Brownian Dynamics simulations}
\titlerunning{Tagged-particle motion in glassy systems under shear}
\date{\today}
\author{Matthias Kr\"uger\thanks{Present address: Department of Physics, Massachusetts Institute of Technology, Cambridge, Massachusetts 02139, USA; Electronic address:kruegerm@mit.edu} \and Fabian Weysser \and Matthias Fuchs}

\institute{Fachbereich Physik, Universit\"at Konstanz,
  78457 Konstanz, Germany}
\abstract{
We study the dynamics of a tagged particle in a glassy system under shear. The recently developed integration through transients approach based on mode coupling theory, is continued to arrive at the equations for the tagged particle correlators and the mean squared displacements. The equations are solved numerically for a two dimensional system, including a nonlinear stability analysis of the glass solution, the so called $\beta$-analysis. We perform Brownian Dynamics simulations in 2-D and compare with theory. After switch on, transient glassy correlation functions show strong fingerprints of the stress overshoot scenario, including, additionally to previously studied superexponential decay, a shoulder-like slowing down after the overshoot. We also find a new type of Taylor dispersion in glassy states which has intriguing similarity to the known low density case. The theory qualitatively captures most features of the simulations with quantitative deviations concerning the shear induced timescales. We attribute these deviations to an underestimation of the overshoot scenario in the theory. 
}


\maketitle

\section{Introduction}\label{sec:introduction}
The motion of a tagged particle, expressed e.g. through its mean squared displacement (MSD), is a well known and very intuitive indicator for the dynamics of a system. For a single Brownian particle (dilute limit) under shear, the MSD is very anisotropic and shows superdiffusive motion for the direction of shear \cite{Elrick62}, an effect called Taylor dispersion. For the shear pointing in $x$-direction with shear rate $\dot\gamma$ and varying in the $y$-direction, the MSDs in the dilute limit for the different directions read (see the precise definitions below),

\begin{subequations}\label{eq:elrick}
\begin{align}
&\left\langle [z(t)-z(0)]^2\right\rangle^{(\dot\gamma)}=\left\langle [y(t)-y(0)]^2\right\rangle^{(\dot\gamma)}=2 \,D_0 \,t,\\
&\left\langle [x(t)-x(0)]^2\right\rangle^{(\dot\gamma)}=2\,D_0\,{t}+y(0)^2\,\dot\gamma^2 t^2+\frac{2}{3}\,D_0\,\dot\gamma^2\,t^3.\\
&\left\langle [x(t)-x(0)][y(t)-y(0)]\right\rangle^{(\dot\gamma)}=D_0\,\dot\gamma\,t^2.
\end{align}
\end{subequations}
Shearing speeds up the random (non-affine) motion along the direction of the flow because fluctuations along the  gradient ($y$-) direction let the particle experience varying solvent flows. Random displacements along the gradient direction therefore increase the displacement fluctuations in flow direction.
At higher densities, the situation is not as clear and has been studied extensively in the past few years in experiments, simulations and theory (mostly in low density expansions \cite{Morris96}). Systems near the glass transition have only been studied in experiments and simulations before \cite{Besseling07,Miyazaki04,Varnik08,Zausch08,Foss99,Eisenmann10}. At high densities, generally, the MSDs for the directions perpendicular to the shear direction have been found diffusive at long times, with diffusivities depending on shear rate in contrast to the single particle case in Eq.~\eqref{eq:elrick}: The shear influence can only be transformed to the directions perpendicular to shear by particle interactions. In \cite{Varnik08}, it has been seen that the MSD for the $x$-direction grows indeed cubically in time, for a system near the glass transition. Nevertheless, the quantitative relation between the different directions has not been demonstrated.

For a system of non-Brownian particles \cite{Sierou04}, where the particles attain diffusive motion for the directions perpendicular to shear only due to interactions, the relations for the different directions are similar to Eq.~\eqref{eq:elrick}. In contrast to Eq.~\eqref{eq:elrick}, the shear dependent diffusivities are anisotropic in general.

For super-cooled liquids in general, the dynamics of the tagged particle (as visualized by the MSD or the incoherent density correlation function) has been shown to exhibit nontrivial features after switch on of shear, connected to the shear stress as function of time \cite{Zausch08,Krueger10b}. After switch on, the stress reaches a maximum (sometimes referred to as static yield stress), where the glass yields, followed by a monotonic decay of the stress down to the stationary value giving the 'flow curve'. This scenario, called 'stress overshoot', was shown to be visible in the transient tagged particle functions, as the MSD is superdiffusive and the density correlation function is superexponential  right after the stress maximum.

In this contribution, we study the tagged-particle motion close to vitrification including shear-melted glasses. We focus on the transient dynamics after switching on the shear, which we analyze by mode coupling theory and in Brownian dynamics simulations. Our paper is composed of the following sections. In section \ref{sec:susc}, we introduce the considered system and present the derivation of the equation of motion for the incoherent density correlation function in section \ref{sec:exeq}. Section \ref{sec:results} discusses its numerical solution in detail, including a $\beta$-analysis and the discussion of master-curves for small shear rates. Section \ref{sec:MSDs} is devoted to derive analytic expressions for the MSDs, discussing the Taylor dispersion near the glass transition. Numerical results are given in section \ref{sec:MSDnum}. Section \ref{sec:MSDstat} closes the theoretical part of the paper by discussing the waiting time dependence of the MSDs after switch on. 

Finally, we show the results of our simulations in section \ref{sec:sim}, which in subsections \ref{sec:Corr}, \ref{sec:simbeta} and \ref{sec:simalpha} presents the density correlation functions, the focus on the dynamics near the critical plateau and the master-curves, respectively. In these subsections, the glassy transient correlators will be shown to have the interesting features of shoulders, which we attribute to the slowing down of the system after the stress-overshoot. Subsection \ref{sec:simMSD} shows the MSDs for the different directions, demonstrating the validity of the relations connecting the different directions as found in section \ref{sec:MSDs}.

\section{Microscopic starting point}\label{sec:susc}
We consider a system of $N$ spherical Brownian (bath-) particles of diameter $d$, and the spherical tagged particle of diameter $d_s$ dispersed in a solvent. The system has volume $V$. The bath particles have bare diffusion constants $D_0$, the tagged particle $D_0^s$. The interparticle force acting on particle $i$ ($i=1,\dots, N,s$) at position $\vct{r}_i$ is given by $\vct{F}_i=-\partial \mathcal{U}(\{\vct{r}_j\})/\partial \vct{r}_i$, where $ \mathcal{U}$ is the total potential energy. We neglect hydrodynamic interactions to keep the description as simple as possible. These are also absent in our computer simulations to which we will compare the results.

The external driving, viz. the shear, acts on the particles via the solvent flow velocity $\vct{v}(\vct{r})=\dot\gamma y \hat{\vct{x}}$, i.e., the flow points in $x$-direction and varies in $y$-direction. $\dot\gamma$ is the shear rate. The particle distribution function $\Psi(\Gamma\equiv\{{\vct{r}_i}\},t)$ obeys the Smoluchowski equation \cite{Dhont,Fuchs05},
\begin{eqnarray}\label{eq:smol}
\partial_t \Psi(\Gamma,t)&=&\Omega \; \Psi(\Gamma,t),\nonumber\\
\Omega&=&\Omega_e+\delta\Omega=\sum_{i=1}^{N,s}\boldsymbol{\partial}_i\cdot\left[\boldsymbol{\partial}_i-{\bf F}_i - \kap\cdot\vct{r}_i\right],
\end{eqnarray}
with $\kap=\dot\gamma\hat{\vct{x}}\hat{\vct{y}}$ for the case of simple shear.
$\Omega$ is called the Smoluchowski operator (SO) and it is built up by the equilibrium SO, $\Omega_e=\sum_{i}\boldsymbol{\partial}_i\cdot[\boldsymbol{\partial}_i-{\bf F}_i]$ of the system without shear and the shear term $\delta\Omega=-\sum_i\boldsymbol{\partial}_i\cdot \kap\cdot\vct{r}_i$. We introduced dimensionless units, where lengths, energy and time are measured in units of  $d$, $k_BT$ and $d^2/D_0$, respectively. The effect of shear relative to Brownian motion is measured by the (bare) Peclet number Pe$_0=\dot\gamma d^2/D_0$, which in these units agrees with the shear rate. 

The formal H-theorem \cite{Risken} states that the system reaches the equilibrium distribution $\Psi_e$ at long times, viz. $\Omega_e \Psi_e=0$, without shear. Under shear, the system reaches the stationary distribution $\Psi_s$ with $\Omega \Psi_s=0$. Ensemble averages in equilibrium and in the stationary state are denoted
\begin{subequations}
\begin{eqnarray}
\left\langle\dots\right\rangle&=&\int d\Gamma \Psi_e(\Gamma) \dots,\\
\left\langle\dots\right\rangle^{(\dot\gamma)}&=&\int d\Gamma \Psi_s(\Gamma) \dots ,
\end{eqnarray}
\end{subequations}
respectively.
\section{Equation of Motion for the Transient Incoherent Correlator}\label{sec:exeq}
The information about the average dynamics of a tagged particle is contained completely in the so called incoherent density correlator. Under shear, one can define different dynamical correlation functions, as discussed in Refs.~\cite{Krueger10b,Zausch08}. We will start in this section with the transient one, for which the external shear is switched on at $t=0$. It is the general strategy in the MCT-ITT approach (an extension of mode coupling theory (MCT) \cite{Goetze91} to sheared systems, where ITT stands for 'integration through transients' \cite{Fuchs05}), to start in deriving the transient quantities.  In the coherent case this is justified by the generalized Green Kubo relations for the stress \cite{Fuchs09} and the fact that the transient correlator can be obtained with the equilibrium structure factor as only input. Here it is  a natural continuation to derive the equation for the transient incoherent correlator, since we will be able to use many insights gained from both the coherent and the equilibrium case. Furthermore, this approach will lead to the stationary mean square displacements (see section \ref{sec:MSDstat}), one of the main goals of this contribution, and the transient incoherent correlator can serve to derive other observables in ITT, in the future.
The  transient incoherent density correlator $\Phi_{\bf q}^{s}(t)$ (the intermediate scattering function) is defined as
\begin{equation}
\Phi_{\bf q}^{s}(t)=\left\langle e^{-i\pr}\ti e^{i\prt}\right\rangle=\left\langle \varrho_{\bf q}^{s*}\ti \varrho_{\vct{q}(t)}^s\right\rangle\label{eq:corrinco},
\end{equation}
with the particle position $\vct{r}_s$. In contrast to the coherent case, the normalization of the correlator is unity since $e^{-i\pr}e^{i\pr}=1$ holds. On the right hand side the advected wavevector, a specialty of the ITT-approach \cite{Fuchs09} appears. It reads 
\begin{equation}
{\bf q}(t)=\vct{q}-\dot\gamma t q_x \vct{e}_y.\label{eq:advwav}
\end{equation}
It appears in Eq.~\eqref{eq:corrinco} because of translational invariance of the infinite system \cite{Fuchs08b}. All wavevectors other then Eq.~\eqref{eq:advwav} lead to zero in Eq.~\eqref{eq:corrinco} \cite{Fuchs05,Fuchs09}. Due to this advection, the density correlator is, strictly speaking, no autocorrelation function for $q_x\not=  0$. It can be rewritten using $e^{-\delta\Omega^\dagger t} e^{i\pr}=e^{i\prt}$, 
\begin{equation}
\Phi_{\bf q}^{s}(t)=\left\langle e^{-i\pr}\ti e^{-\delta\Omega^\dagger t} e^{i\pr}\right\rangle.
\end{equation}
We see that $\Phi_{\bf q}^{s}(t)$ is an autocorrelation function with respect to the time evolution of 
\begin{equation}
U(t)=\ti e^{-\delta\Omega^\dagger t}=e^{\Omega^\dagger_e t+\delta\Omega^\dagger t}e^{-\delta\Omega^\dagger t}.
\end{equation}
It is worth noting that, if $\Omega^\dagger_e$ and $\delta\Omega^\dagger$ commuted, we would have $\Phi_{\bf q}^{s}(t)=\Phi^{s(e)}_{q}(t)$, the equilibrium correlator. This is, of course, not the case. The following derivation of the equation of motion for $\Phi_{\bf q}^{s}(t)$ is analogous to the coherent case \cite{Fuchs09} and we will therefore be very brief. 

The time dependence of the evolution operator $U(t)$ can be found by differentiation,
\begin{equation}
\partial_t U(t)= \ti(\Omega^\dagger-\delta\Omega^\dagger) e^{-\delta\Omega^\dagger t}=\ti\,\Omega^\dagger_e \,e^{-\delta\Omega^\dagger t}\label{eq:U}.
\end{equation}
We see that the equilibrium operator appears. To proceed, it is reasonable to define a Hermitian operator as was suggested in Ref.~\cite{Fuchs09},
\begin{equation}
\Omega^\dagger_a(t)=e^{\overline{\delta\Omega^\dagger}t} \Omega^\dagger_e e^{-\delta\Omega^\dagger t},
\end{equation} 
with $\overline{\delta\Omega^\dagger}=\sum_i\vct{r}_i\cdot\kap^{T}\cdot(\boldsymbol{\partial}_i+\vct{F}_i)$. $\overline{\delta\Omega^\dagger}$ is the adjoined of $-{\delta\Omega^\dagger}$ in the equilibrium average. It follows that $\Omega^\dagger_a(t)$ is Hermitian in the equilibrium average, because $\Omega^\dagger_e$ is Hermitian in this average \cite{Fuchs05},
\begin{align}
\left\langle f^* \Omega^\dagger_a(t) g\right\rangle&=\left\langle (e^{-\delta\Omega^\dagger t} f^*) \Omega^\dagger_e e^{-\delta\Omega^\dagger t} g\right\rangle\\&=\left\langle (\Omega^\dagger_e e^{-\delta\Omega^\dagger t} f^*) e^{-\delta\Omega^\dagger t} g\right\rangle=\left\langle g^*\, \Omega^\dagger_a(t) f\right\rangle^*.
\end{align}
And with $f=g$ the above equation also shows that the time dependent eigenvalues of $\Omega^\dagger_a(t)$ are real and negative. Because of       
\begin{equation}
\langle \varrho_{\vct{q}}^{s*}e^{\overline{\delta\Omega^\dagger}t}=\langle e^{-i\pr}e^{\overline{\delta\Omega^\dagger}t}=\langle e^{-i\prt}=\langle \varrho_{\vct{q}(t)}^{s*},
\end{equation} 
$\Omega^\dagger_a(t)$  has identical matrix elements as the equilibrium operator for the case of density fluctuations, only the  densities are replaced by their time dependent analogs as we will see when regarding the initial decay rate in Eq.~\eqref{eq:spec}. 

The equations of motion are derived in the spirit of the Zwanzig-Mori projection operator formalism \cite{Zwanzig}, where we use the time dependent single particle density projector 
\begin{equation}
P^s(t)=\sum_{\bf q} \varrho_{\vct{q}(t)}^s \rangle\langle \varrho_{\vct{q}(t)}^{s*}
\end{equation}  
with complement $Q^s(t)=1-P^s(t)$. We abbreviate $P^s(0)=P^s$, the well known single particle projector used for the quiescent system \cite{Goetze91}. With this, Eq.~\eqref{eq:U} can be rewritten such that the well behaved operator appears
\begin{align}
\partial_t U(t)&= \ti\,(P^s(t)+Q^s(t))\Omega^\dagger_e \,e^{-\delta\Omega^\dagger t}\notag\\&=U(t)\,(P^s\Omega^\dagger_a(t)+\Omega_r^\dagger(t)),
\end{align}
with
\begin{align}
\Omega_r^\dagger(t)&=e^{\delta\Omega^\dagger t}Q^s(t)\Omega_e^\dagger e^{-\delta\Omega^\dagger t}.
\end{align}
At $\dot\gamma=0$, $\Omega^\dagger_r(t)$ is perpendicular to density fluctuations, which is not the case for $\dot\gamma\not=0$. The part which is not perpendicular can be split off by writing
\begin{equation}
\Omega^\dagger_r(t)=\Omega^\dagger_Q(t)+\Omega^\dagger_\Sigma(t).
 \end{equation}
The first part is perpendicular to density fluctuations, $P^s\Omega^\dagger_Q(t)=0$, while the other one is not. The two parts read
\begin{subequations}
\begin{eqnarray}
\Omega^\dagger_Q(t)&=&e^{\overline{\delta\Omega^\dagger} t}Q^s(t)\Omega_e^\dagger e^{-\delta\Omega^\dagger t},\\
\Omega^\dagger_\Sigma(t)&=&e^{\overline{\delta\Omega^\dagger} t}\Sigma(t)\,Q^s(t)\,\Omega_e^\dagger e^{-\delta\Omega^\dagger t},\label{eq:omsig}
\end{eqnarray}
\end{subequations}
with the function $\Sigma(t)$ given by \cite{Fuchs09}($\sigma_{xy}=-\sum_iF_i^x y_i$)
\begin{equation}
\Sigma(t)=\dot\gamma\int_0^t\!\!dt' e^{-\overline{\delta\Omega^\dagger} t'}\sigma_{xy} e^{\delta\Omega^\dagger t'}.\label{eq:Sigma}
\end{equation}
Because of $\Sigma(t)$, the second part of $\Omega^\dagger_r(t)$ couples to density fluctuations.

As is done in the equilibrium case, a reduced time evolution operator is employed which satisfies \begin{equation}
\partial_t \,U_r(t,t')=U_r(t,t')\,\Omega_r^\dagger (t).
\end{equation}
Its formal solution is given in terms of a time ordered exponential, where operators are ordered from left to right as time increases \cite{Kawasaki73},
\begin{equation}
U_r(t,t')=e_-^{\int_{t'}^t ds \Omega_r^\dagger(s)}.
\end{equation}
We still need the connection between reduced and full evolution operators given by
\begin{equation}
U(t)=U_r(t,0)+\int_0^t dt' U(t') P^s \Omega^\dagger_a(t') U_r(t,t').
\end{equation}
Taking its time derivative leads to the useful operator relation,
\begin{align}
\partial_t U(t)&= U_r(t,0)\Omega_r^\dagger(t)+U(t)P^s\Omega^\dagger_a(t)\nonumber\\&+\int_0^t dt'U(t')P^s \Omega^\dagger_a(t')U_r(t,t')\Omega^\dagger_r(t). 
\end{align}
The equation of motion for the desired correlator now follows by sandwiching the expressions above with single particle density fluctuations $e^{i\pr}$. As already noted, the operator $\Omega_r^\dagger(t)$ is not perpendicular to these density fluctuations and the first term on the right hand side does not vanish as it does at $\dot\gamma=0$. The equation of motion hence contains an extra term,
\begin{equation}
\partial_t \Phi_{\bf q}^{s}(t)+\Gamma^s_{\bf q}(t)\Phi_{\bf q}^{s}(t)+\int_0^tdt' M^{s}_{\bf q}(t,t')\,\Phi_{\bf q}^{s}(t')=\Delta^{s}_{\bf q}(t). \label{eq:MCT1}
\end{equation}
The extra term $\Delta^{s}_{\bf q}(t)$ reads 
\begin{equation}
\Delta^{s}_{\bf q}(t)=\left\langle e^{-i\pr}U_r(t,0)\Omega_r^\dagger(t)e^{i\pr} \right\rangle,\label{eq:deltaterm}
\end{equation}
it vanishes only at time $t=0$ and grows to lowest order like $\dot\gamma t$. An analogous term appears in the equation of motion for the coherent correlator in Ref.~\cite{Fuchs09}. As argued there, its appearance  is the only disadvantage of this approach compared to an earlier one (Ref.~\cite{Fuchs05}). In contrast to Ref.~\cite{Fuchs05}, the initial decay rate is positive; it is equal to the equilibrium initial decay rate for advected wavevectors (recall that $\Omega^\dagger_a$ has negative semi-definite spectrum),
\begin{align}
\Gamma^s_{\bf q}(t)&=\Gamma^{s(e)}_{\vct{q}(t)}\notag=-\left\langle e^{-i\pr} \Omega_a^\dagger \,e^{i\pr}\right\rangle\\&=-\left\langle e^{-i\prt} \Omega_e^\dagger \,e^{i\prt}\right\rangle=q^2(t)\geq 0.\label{eq:spec}
\end{align}
The positivity of the initial decay rate makes the numerical analysis of the equations  below more stable. 
The memory function $M^{s}_{\bf q}(t)$ contains on the left hand side the well behaved operator $\Omega_a^\dagger$,
\begin{equation}
M^{s}_{\bf q}(t,t')=-\left\langle\varrho_{\bf q}^{s*}\Omega^\dagger_a(t')U_r(t,t')\Omega^\dagger_r(t) \varrho_{\bf q}^s\right\rangle\label{eq:MMCT}.
\end{equation}
If we knew an approximation for $M^{s}_{\bf q}(t,t')$ in terms of the correlator itself, the equation would be closed apart from $\Delta^{s}_{\bf q}(t)$.
But MCT approximations for Eq.~\eqref{eq:MMCT} are not desirable,  as was discussed in Refs.~\cite{Fuchs05,Fuchs09}: Approximating $M^{s}_{\bf q}(t,t')$ in Eq.~\eqref{eq:MCT1}, one would have to be very careful to obtain an equation which describes slow dynamics. This is not the case for Eq.~\eqref{eq:incoexact} below. Because of this, we perform a second projection step following Ref.~\cite{Fuchs09}. To decompose the reduced SO appearing in $U_r(t,t')$, we use the projector
\begin{equation}
\tilde P^s(t)=\varrho_{\bf q}^s\rangle \frac{1}{\langle{\varrho_{\bf q}^{s*}}\Omega^\dagger_a(t)\varrho_{\bf q}^s\rangle}\langle{\varrho_{\bf q}^{s*}}\Omega^\dagger_a(t),
\end{equation}
with complement $\tilde Q^s(t)$. While $\tilde P^s(t)$ is, strictly speaking, not a projector because it is not Hermitian, it is still idempotent, $\tilde P^s(t)\tilde P^s(t)=\tilde P^s(t)$.
It is applied in the following way,
\begin{align}
\Omega^\dagger_r(t)&=\Omega^\dagger_r(t)(\tilde Q^s(t)+\tilde P^s(t))\notag\\&=\Omega^\dagger_i(t)+\Omega_r^\dagger(t)\varrho_{\bf q}^s\rangle \frac{1}{\langle{\varrho_{\bf q}^{s*}}\Omega^\dagger_a(t)\varrho_{\bf q}^s\rangle}\langle{\varrho_{\bf q}^{s*}}\Omega^\dagger_a(t).
\end{align} 
One can then relate $M^{s}_{\bf q}(t,t')$ to another memory function, $m^{s}_{\bf q}(t,t')$, which is governed by the irreducible operator $\Omega_{i}^\dagger(t)$ \cite{Kawasaki95,Cichocki87}. The lengthy calculation which leads to the equations below is presented in detail in Ref.~\cite{Fuchs09}: The equation of motion can then (with the use of the theory of Volterra integral equations \cite{Tricomi}) be written as
\begin{align}
\partial_t\Phi_{\bf q}^{s}(t)+\Gamma^{s}_{\bf q}(t)\left\{\Phi_{\bf q}^{s}(t)+\int_0^t dt' m^{s}_{\bf q}(t,t') \partial_{t'}\Phi_{\bf q}^{s}(t')\right\}\notag\\=\tilde\Delta^{s}_{\bf q} (t),\label{eq:incoexact}
\end{align}
with the new memory function 
\begin{equation}
m^{s}_{\bf q}(t,t')=\frac{1}{\Gamma_{\bf q}(t)\Gamma_{\bf q}(t')}\left\langle{\varrho_{\bf q}^s}^*\Omega^\dagger_a(t')U_i(t,t')\Omega^\dagger_r(t) \varrho_{\bf q}^s\right\rangle.
\end{equation}
It is governed by the irreducible operator,
\begin{equation}
U_i(t,t')=e_{-}^{\int_{t'}^t ds \,\Omega_i^\dagger(s)}.
\end{equation}
Eq.~\eqref{eq:incoexact} has an extra term compared to the familiar one known from quiescent MCT \cite{Fuchs98,chong2009}: The term on the right hand side arose from $\Delta^{s}_{\bf q}(t)$ in Eq.~\eqref{eq:MCT1},
\begin{equation}
\tilde\Delta^{s}_{\bf q}(t)=\langle {\varrho_{{\bf q}}^{s*}}U_i(t,0)\Omega^\dagger_r(t)\varrho_{{\bf q}}^s\rangle.
\end{equation}
It also vanishes at $t=0$ and grows in leading order like $\dot\gamma t$. It does hence not influence the fast decay onto the plateau for $\dot\gamma\to0$.
This exact set of equations for the incoherent transient density correlator is now suitable for approximations in order to get a closed equation for $\Phi^{s}_{\bf q}(t)$.

The first simplification concerns the source term $\tilde\Delta^{s}_{\bf q}(t)$ arising from the stress expression $\Sigma(t)$ in Eq.~\eqref{eq:Sigma}. In Ref.~\cite{Fuchs09} it is suggested to set $\Sigma(t)\equiv0$ in leading approximation. This leads immediately to $\tilde\Delta^{s}_{\bf q}(t)\equiv 0$ since $\Omega^\dagger_\Sigma(t)=0$ follows in Eq.~\eqref{eq:omsig}, and with it $\Delta^{s}_{\bf q}(t)=0$ in Eq.~\eqref{eq:deltaterm}. We note the identity
\begin{equation}
e^{\delta \Omega^\dagger t}=e^{\overline{\delta \Omega^\dagger}t} (1+\Sigma(t)),
\end{equation} 
and hence  $e^{\delta \Omega^\dagger t}=e^{\overline{\delta \Omega^\dagger}t}$ with $\Sigma(t)\equiv0$. 
Approximating $\Sigma(t)\equiv0$ leads also to a simplification of the memory function $m^{s}_{\bf q}(t,t')$ because $\Omega^\dagger_r$ reduces to $\Omega_Q^\dagger(t)$. With this, the time evolution $U_i(t,t')$ becomes 
\begin{equation}
U_i^Q(t,t')=e_{-}^{\int_{t'}^t ds \,\Omega^\dagger_Q(s) \tilde Q^s(s)}.
\end{equation}
It is finally in the space perpendicular to density fluctuations, $P^sU_i^Q(t,t')=0=U_i^Q(t,t')P^s$. For the memory function follows 
\begin{align}
&q^2(t)q^2(t') m^{s}_{\bf q}(t,t')\notag\\\notag&=\left\langle\varrho_{{\bf q}(t')}^{s*}\Omega^\dagger_e e^{-\delta\Omega^\dagger t'}U_i(t,t') e^{\delta\Omega^\dagger t} Q^s(t)\Omega^\dagger_e \varrho_{{\bf q}(t)}^s\right\rangle,\\
&=\left\langle\varrho_{{\bf q}(t')}^{s*}\Omega^\dagger_e\,Q^s(t')\, e^{-\overline{\delta\Omega^\dagger} t'}U_i^Q(t,t') e^{\delta\Omega^\dagger t} Q^s(t)\Omega^\dagger_e \varrho_{{\bf q}(t)}^s\right\rangle.\label{eq:memorysimpl}
\end{align}
The allowed insertion of $Q^s(t')$ on the left hand side can easily be verified; inserting $P^s(t')$ at the same position leads to zero.
For the following mode coupling approximations, the pair density projector is used, which is assumed to describe the slow dynamics in the glassy regime. In contrast to the coherent case, the pair projector in the incoherent case consists of the product of coherent and incoherent fluctuations \cite{Goetze91}. This has a physical reason; the fluctuating force $Q_s\Omega^\dagger \varrho^s_{\bf q}=  Q_s (i{\bf q}\cdot {\bf F}_s e^{i{\bf q}\cdot{\bf r}_s})$ on the tagged particle depends on the tagged particle and the collective dynamics, i.e., the dynamics of the surrounding bath particles. Technically this is achieved by the projector,
\begin{equation}
P^s_2(t)=\sum_{\vct{p},\vct{k}}\frac{\varrho_{{\bf p}(t)}^s\varrho_{{\bf k}(t)}\rangle\langle\varrho^{s*}_{\vct{p}(t)}\varrho^*_{\vct{k}(t)}}{N S_{k(t)}}, \label{eq:pairden}
\end{equation}
with $\varrho_{\bf k}=\sum_i^N e^{i{\bf k}\cdot{\bf r}_i}$ the density of the bath particles and $S_k=\langle \varrho^*_{\bf k} \varrho_{\bf k}\rangle/N$ the structure factor.
Note that in contrast to the coherent pair projector, the two densities can be distinguished here and the wavevectors are not ordered. No counting factor will appear.
The memory function \eqref{eq:memorysimpl} is written,
\begin{align}
m^{s}_{\bf q}(t,t')\approx\frac{1}{q^2(t)q^2(t')}\bigl\langle\varrho_{{\bf q}(t')}^{s*}\Omega^\dagger_e\,Q^s(t')P^s_2(t')\, e^{-\overline{\delta\Omega^\dagger} t'}\notag\\U_i^Q(t,t') e^{\delta\Omega^\dagger t} P^s_2(t)\,Q^s(t)\Omega^\dagger_e \varrho_{{\bf q}(t)}^s\bigr\rangle,
\end{align}
and in accordance with Ref.~\cite{Fuchs09}, the appearing four point correlation function is approximated as the product of correlators with full dynamics,
\begin{eqnarray}
\left\langle\varrho^{s*}_{\vct{p}(t')}\varrho^*_{\vct{k}(t')}e^{-\overline{\delta\Omega^\dagger} t'}U_i^Q(t,t') e^{\delta\Omega^\dagger t}\varrho_{{\bf p'}(t)}^s\varrho_{{\bf k'}(t)}\right\rangle\nonumber\\
\approx NS_{k(t')} \Phi_{{\bf p}(t')}^{s}(t-t')\,\Phi_{{\bf k}(t')}(t-t')\,\delta_{\vct{p},\vct{p'}}\delta_{\vct{k},\vct{k'}}.
\end{eqnarray}
This factorization of the four point function is the major approximation  in this approach. A similar approximation is used also in quiescent MCT \cite{Goetze91}.
The remaining parts of the vertex are now found easily, since they are identical as in equilibrium using the advected wavevectors instead of the time independent ones. The vertex in equilibrium reads (we have already inserted the restriction of $\vct{p}=\vct{k}-\vct{q}$),
\begin{equation}
V_{\vct{q}\vct{k}}=\frac{\left\langle\varrho^{s}_{\bf k-q}\varrho_{\bf -k}\,Q^s\,\Omega^\dagger_e\varrho_{\bf q}^s\right\rangle}{N S_k}=\frac{1}{V}\vct{k}\cdot\vct{q}\,c_k^{s},\label{eq:vertexinco}
\end{equation}
where $c_q^{s}=\langle \varrho^{s*}_{q}\varrho_{q}\rangle/(n S_q)$ is the direct single particle correlation function \cite{Hansen}, $n=N/V$ is the density. The sum over bath particles does not contain the tagged particle, and we have $c_q^{s}=(S_q-1)/(n S_q)$ if the tagged particle is identical to the bath particles. Summarizing, we find the following approximate equation of motion for the incoherent transient density correlator, 
\begin{equation}
\partial_t\Phi_{\bf q}^{s}(t)+\Gamma^{s}_{\bf q}(t)\left\{\Phi_{\bf q}^{s}(t)+\int_0^t dt' m^{s}_{\bf q}(t,t') \partial_{t'}\Phi_{\bf q}^{s}(t')\right\}=0,\label{eq:incofin}
\end{equation}
with $\Gamma^{s}_{\bf q}(t)=q^2(t)$ (compare Eq.~\eqref{eq:spec}) and 
\begin{align}
&m^{s}_{\bf q}(t,t')\approx\frac{1}{N}\sum_{\bf k}\frac{\vct{k}(t)\cdot\vct{q}(t)}{q^2(t)}\frac{\vct{k}(t')\cdot\vct{q}(t')}{q^2(t')}\notag\\
&n^2 c_{k(t)}^{s}c_{k(t')}^{s} S_{k(t')}\,\Phi_{\vct{k}(t')-\vct{q}(t')}^{s}(t-t')\Phi_{\vct{k}(t')}(t-t')\nonumber.
\end{align}
Changing the summation index from $\vct{k}$ to $\vct{k}'=\vct{k}(t')$ (and immediately renaming the dummy variable from $\vct{k'}$ to $\vct{k}$), we get 
\begin{align}
&m^{s}_{\bf q}(t,t')=\frac{1}{N}\sum_{{\bf k}}\frac{\vct{k}(t-t')\cdot\vct{q}(t)}{q^2(t)}\frac{\vct{k}\cdot\vct{q}(t')}{q^2(t')}\notag\\
&n^2 c_{k(t-t')}^{s}c_{k}^{s} S_{k}\,\Phi_{\vct{k}-\vct{q(t')}}^{s}(t-t')\Phi_{\vct{k}}(t-t')\label{eq:mfin}.
\end{align}
We see that this final form only depends explicitly on $t'$ via $\vct{q}(t)$ since we can use, e.g. ${\bf q}(t)={\bf q}(t')(t-t')$ to write $m^{s}_{\bf q}(t,t')=\bar{m}^s_{\vct{q}(t')}(t-t')$ with   
\begin{align}
&\bar{m}^s_{\vct{q}}(t-t')=\frac{1}{N}\sum_{{\bf k}}\frac{\vct{k}(t-t')\cdot\vct{q}(t-t')}{q^2(t-t')}\frac{\vct{k}\cdot\vct{q}}{q^2}\notag\\
&n^2 c_{k(t-t')}^{s}c_{k}^{s} S_{k}\,\Phi_{\vct{k}-\vct{q}}^{s}(t-t')\Phi_{\vct{k}}(t-t')\label{eq:mfin2}.
\end{align}
Through the pair density projector, the dynamics of the incoherent correlator is coupled to the coherent correlator. Eq.~\eqref{eq:incofin} can therefore only be solved if the corresponding equation for the coherent dynamics has been solved before. This coupling is physically intuitive, since a (large enough) tagged particle can only move if the surrounding particles move. There is a certain percolation threshold for the size of the tagged particle, below which it is mobile even if the bath is arrested \cite{Franosch94}. Yet, we will in the numerical solutions consider the case where the tagged particle is one of the bath particles, (i.e., the tagged particle is much larger then the percolation threshold). Then, at $\dot\gamma=0$, the dynamics of the tagged particle follows the dynamics of the bath particles \cite{Fuchs98,Goetze91,Goetze95}, i.e., the tagged particle is trapped if, and only if, the bath is arrested. 

The memory function \eqref{eq:mfin} depends on $t'$ and $t-t'$. This complicates the following analysis because the convolution theorem cannot be applied. It probably originates from the fact that we investigate the transient regime which is not time translationally invariant. An equation for the stationary correlator should contain a memory function depending on $t-t'$ only.
\section{Results for the transient incoherent correlator}\label{sec:results}
\subsection{Numerical details}
Let us turn to the numerical evaluation of Eq.~\eqref{eq:incofin} which we performed in $D=2$ dimensions for a system of equal sized hard discs ($d_s=d$). The only thermodynamic control parameter is the area fraction $\eta=\frac{\pi N}{4V}$. 

The solution for $D=3$ is as yet numerically too costly in computer time and memory. For $D=2$, we used a spherical grid with 100 points in radial direction, $q=0.2,0.6,1.0,\dots, 38.8$. The angular space was divided in $96$ portions, giving a grid of $\theta=0.065,0.13,\dots, 2\pi$. The number $96$ is often divisible by $2$ allowing us to give the correlator for angles $\theta=\pi/2,\pi/4,\pi/8$ and so on, which are the most interesting to be analyzed. 
Note that this grid is different compared to the one used in \cite{fabian_proc}, where (only) the coherent density correlators were determined. While the resulting solutions are very similar, the current grid has the advantage that the correlators for constant $q$ can be given for all $\theta$, so anisotropy effects can be well studied. This is not possible for the Cartesian grid used in \cite{fabian_proc}. On the other hand, the numerical algorithm for the spherical grid involves more interpolation procedures, since the vector $\bf q-k$ is not on a grid-point.

From our discretization follows the critical packing of $\eta_c=0.6985658$ and the exponent parameter $\lambda=0.7155$. The latter determines all power-law exponents of the theory. These values differ slightly from the ones found in Ref.~\cite{Bayer07} ($\eta_c=0.696810890$ and $\lambda=0.7167$) due to the different discretization of $q$-space, which is finer in Ref.~\cite{Bayer07}.
\subsection{Correlator $\Phi^s_{\bf q}$}\label{sec:corr}
As noted above, Eqs.~\eqref{eq:incofin} and \eqref{eq:mfin} (together with the coherent analogues \cite{Fuchs09}) show the well known bifurcation scenario connected to the glass transition at $\eta_c$, separating the control parameter region with intrinsically ergodic correlators from the one where the correlators only because of flow decay to zero at long times. 

\begin{figure}
\includegraphics[angle=270,width=0.95\linewidth]{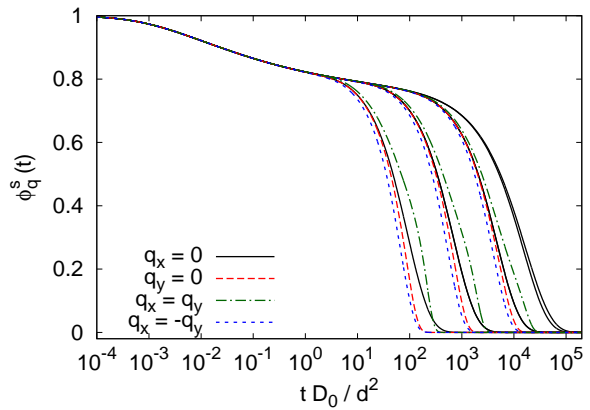}
\caption{\label{fig:incohfl} Transient incoherent density correlator for $\varepsilon=-10^{-3}$ (liquid) and ${ q}d=6.6$. Shear rates are $\dot\gamma=10^{-n}$ with $n=2, \dots, 6$. For the three largest shear rates, we show the four characteristic directions, for the small rates, only  $q_x=0$ is shown for visibility. The curves for the two smallest rates coincide.}
\end{figure}

While this transition is a cooperative effect, i.e., it happens for all wavevectors $q$ at the same density, the shape of $\phis$ (for both with and without shear) depends on $\bf q$. For densities below the glass transition, i.e. $\varepsilon\equiv\frac{\eta-\eta_c}{\eta_c}<0$, the correlator for the system without shear decays to zero with time scale $\tau_\alpha$, the so called $\alpha$-relaxation time \cite{Goetze91}. The effect of shear does then depend on the dressed Peclet or Weissenberg number Pe$=\dot\gamma\tau_\alpha$. For small shear rates, the effect vanishes,
\begin{equation}
\lim_{\tau_\alpha\dot\gamma\to0}\phis\to\phise,  \hspace{1cm}  \rm liquid.
\end{equation}
This is demonstrated in Fig.~\ref{fig:incohfl} where the correlator for a liquid state ($\varepsilon=-10^{-3}$) is shown at different shear rates. For large Pe, the final decay is dominated by shear, and the correlator is anisotropic in $\bf q$-space, whereas the curve with the smallest shear rate shown ($Pe_0=10^{-6}$) is indistinguishable from the equilibrium curve and the correlator is isotropic here. 

Above or at the critical density, the correlator of the system without shear stays on the plateau characterized by the non-ergodicity parameter $f_q^s$,
\begin{equation}
\lim_{t\to\infty}\phise=f^s_q>0, \hspace{1cm} \rm glass.\label{eq:noner}
\end{equation}
At the transition, $f_q^s$ jumps discontinuously from zero to a finite value, given the size of the tagged particle is not too close to the percolation threshold \cite{Goetze91}. The system under shear, however, is always ergodic, since shear melts the glass, and $\phis$ decays to zero for any finite $\dot\gamma$. Since glassy systems are frozen in without shear, the final decay from the plateau to zero is governed solely by shear, for arbitrarily small $\dot\gamma\to 0$. The dressed Peclet number is always infinite because the intrinsic $\tau_\alpha$ is formally infinite.

\begin{figure}
\includegraphics[angle=270,width=0.95\linewidth]{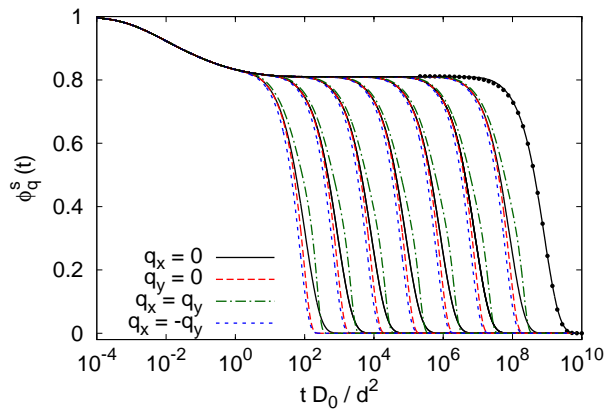}
\caption{\label{fig:incohgl} Transient incoherent density correlator for $\varepsilon=10^{-3}$ (glass) and ${ q}d=6.6$. Shear rates are $\dot\gamma=10^{-n}$ with $n=2, \dots, 9$. For $\dot\gamma=10^{-9}$, we only show the curve for $q_x=0$, together with a fitted compressed exponential with exponent $\mu=1.05$ (dots), see Eq.~\eqref{eq:fit}.}
\end{figure}

Fig.~\ref{fig:incohgl} shows the correlator for a glassy state ($\varepsilon=10^{-3}$) at different shear rates. It is seen that the effect of shear, and the anisotropy in $\bf q$-space, prevails up to arbitrary small $\dot\gamma$. 
For $\varepsilon\geq0$, the functions approach a master function for $\dot\gamma\to0$ and $\dot\gamma t=const.$, which depends only on $\dot\gamma t$. This can be seen in Fig.~\ref{fig:incohgl} and will be discussed in more detail in sec.~\ref{sec:alpha}. For the range of shear rates shown in Fig.~\ref{fig:incohgl}, the anisotropy depends hardly on the shear rate, probably because even $\dot\gamma=10^{-2}$ is already quite well described by the $\dot\gamma\to0$ master function.  
\subsection{$\beta$-Analysis}\label{sec:beta}
Further insight into the dynamics near the critical plateau can be gained by the so called $\beta$-analysis. It is a non-linear stability analysis of the frozen-in structure and consists of an expansion of the equation of motion, Eq.~\eqref{eq:incofin}, around the critical plateau value $f^{sc}_q=f_q^s(\varepsilon=0)$ \cite{Goetze91} defined in Eq.~\eqref{eq:noner}. 
Ref.~\cite{Fuchs03} presents this analysis for the coherent transient density correlator $\Phi_{\bf q}(t)=\langle \rho_{\bf q}^* e^{\Omega^\dagger t} \rho_{\vct{q}(t)}\rangle/\langle \rho_{\bf q}^* \rho_{\bf q}\rangle$, with $\rho_{\bf q}=\sum_ie^{i\vct{q}\cdot\vct{r}_i}$, which can be written near the critical plateau as
\begin{equation}
\Phi_{\bf q}(t)=f_q^c+h_q \label{eq:rec}\left(\mathcal{G}(t)+\mathcal{G}_{\bf q}^{(aniso)}(t)\right)+{\cal O}(\varepsilon).
\end{equation}
$h_q$ is called the critical amplitude. The dynamics near the critical plateau is given by a $q$-independent isotropic part, $\mathcal{G}(t)$, and an anisotropic part, $\mathcal{G}_{\bf q}^{(aniso)}(t)$. The equation of motion of the former is referred to as the $\beta$-equation \cite{Fuchs03}, 
\begin{equation}
\tilde\varepsilon-c^{(\dot\gamma)}(\dot\gamma t)^2+\lambda \mathcal{G}^2(t)=\frac{d}{d t} \int_0^t dt' \mathcal{G}(t-t')\mathcal{G}(t').\label{eq:beta}
\end{equation}
Where $\tilde\varepsilon=C\varepsilon=C(\eta-\eta_c)/\eta_c$ with $C\approx 2.1$ describes the distance from the transition point \cite{Fuchs08b}. For our grid, we find $\lambda\approx0.7155$ and $c^{(\dot\gamma)}\approx 3.4$, see e.g. Ref.~\cite{Fuchs03} for the definitions of these quantities. Note that Eq.~\eqref{eq:beta} is nonlinear (quadratic) at the critical point. This is explained in detail in Ref.~\cite{Goetze91}. The short time behavior of $\mathcal{G}(t)$ must be matched to the short time dynamics of the correlator, $\mathcal{G}(t\to0)=(t_0/t)^a$, where the matching time $t_0$ is determined by the coherent initial decay rate. The critical exponent $a$ obeys (with the $\Gamma$-function) $\lambda=\Gamma^2(1-a)/\Gamma(1-2a)$. From Eq.~\eqref{eq:beta}, we see that the $\beta$-correlator is of order $\sqrt{\varepsilon}$ and $|\dot\gamma t|$, and we keep our discussion to these orders. See Refs.~\cite{Fuchs03,Hajnal09} for more details on the two parameter scaling relation for $\dot\gamma t$ and $\varepsilon$.
The $\beta$-correlator takes for $\varepsilon\geq0$ the solution for long times \cite{Fuchs03},
\begin{equation}
\mathcal{G}(t\gg t_b)=-\sqrt{\frac{c^{(\dot\gamma)}}{\lambda-\frac{1}{2}}}|\dot\gamma| t\equiv-\tilde t.\label{eq:eps=0}
\end{equation}
Eq.~\eqref{eq:eps=0} describes the initialization of  the final shear induced decay from the plateau to zero. One has $t_b=\sqrt{\varepsilon}/|\dot\gamma|$ for $\varepsilon>0$ and $t_b=t_0$  for $\varepsilon=0$.
The shear independent decay from the plateau for the liquid case can be found in Refs.~\cite{Goetze91,Franosch97,Goetze85}.

The anisotropic term in Eq.~\eqref{eq:rec} has been overlooked in Ref.~\cite{Fuchs03}. Since the $\beta$-analysis for the {\it incoherent} transient correlator depends on the coherent one (isotropic and anisotropic), we will only discuss the results here. The detailed derivation of both coherent and incoherent terms will be presented in a forthcoming paper. 

We  consider the case of $\varepsilon\geq0$, because for $\varepsilon<0$, the dynamics is independent of shear for $\dot\gamma\to 0$ and the equilibrium discussion is recovered \cite{Fuchs98}. 
Expanding the incoherent correlator near the critical plateau (for $0\leq\varepsilon\ll1 $ and  $\dot\gamma t\ll1$), we find that the $\beta$-correlator contains an isotropic part given, as in the coherent case, by $\mathcal{G}(t)$, as well as an anisotropic part $\mathcal{G}_{\bf q}^{(s,aniso)}(t)$,
\begin{equation}
\phis=f_q^{sc}+h^{s}_q \left(\mathcal{G}(t)+\mathcal{G}_{\bf q}^{(s,aniso)}(t)\right)+{\cal O}(\varepsilon) \label{eq:betafin}.
\end{equation}
The critical amplitude $h^{s}_q$ is equal to the one at $\dot\gamma=0$ \cite{Fuchs98}. The anisotropic term comes from the lowest order terms in $\dot\gamma t$ of the memory function $m_{\bf q}^s(t,0)$. Here, $\mathcal{G}_{\bf q}^{(aniso)}(t)$, the anisotropic part of the coherent $\beta$-correlator contributes. We find that $\mathcal{G}_{\bf q}^{(s,aniso)}(t)$ is linear in $\dot\gamma t$ and proportional to $q_xq_y$,
\begin{equation}
\mathcal{G}^{(s,aniso)}_{\bf q} (t)=\alpha(|{\bf q}|) \frac{q_xq_y}{q^2}\,\dot\gamma t+\mathcal{O}(\dot\gamma t)^2.\label{eq:alphafunc}
\end{equation} 
The term $q_xq_y$ represents the  
 expected ``quadrupole''-de\-pen\-dence. 
For $q_xq_y>0$, the dynamics is slightly slower than on average and for $q_xq_y<0$ it is slightly faster, i.e., $\alpha(|{\bf q}|)>0$ holds for all $|{\bf q}|$.  
\begin{figure}
\begin{center}
\includegraphics[width=0.95\linewidth]{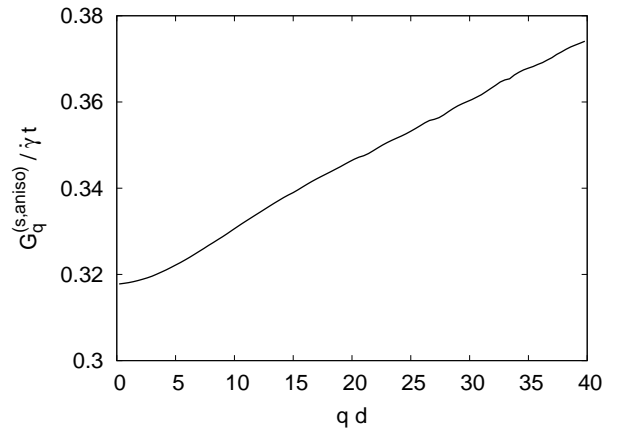}
\caption{\label{fig:vonq2} $\mathcal{G}^{(s,aniso)}_{\bf q} (t)/\dot\gamma t$ at $q_x=q_y$ (identical to $\frac{1}{2}\alpha(|{\bf q}|)$) as function of $|{\bf q}|$.
}
\end{center}
\end{figure}
The function $\alpha(|{\bf q}|)$  increases slowly with $q$, see Fig.~\ref{fig:vonq2}.  
The maximum value  of the anisotropic part (on our grid) is at roughly $0.37 \dot\gamma t$. Still, it renders the slope of $\Phi^s_{\bf q}$ positive for the region $q_x\approx q_y$, since the isotropic contribution $\mathcal{G}(t)$ is initially proportional to $(\dot\gamma t)^2$ (before Eq.~\eqref{eq:eps=0} holds). 

The fact that the anisotropic part is in lowest order proportional to $q_xq_y$ is not unexpected. There are other examples where such a term emerges, e.g. in the distortion of the structure factor under shear; it is in linear order in shear rate also proportional to $q_xq_y$ \cite{Henrich07,Bergenholtz02,Vermant05} for liquid states.

\begin{figure}
\includegraphics[width=1\linewidth]{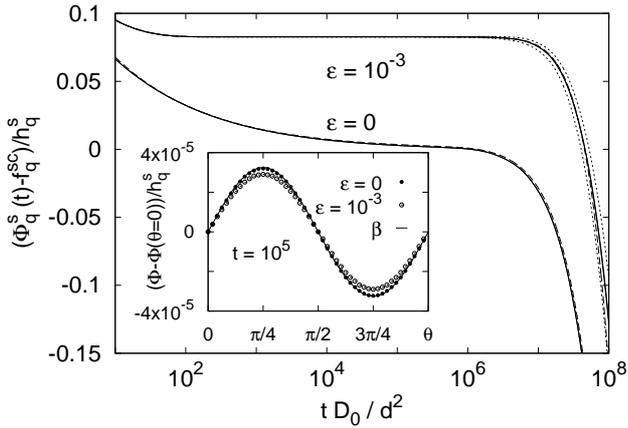}
\caption{\label{fig:fcbetaincoh}
The $\beta$-correlator for the incoherent case. 
We show two glassy states ($\varepsilon=10^{-3}$ and $\varepsilon=0$) with $\dot\gamma=10^{-9}$. The wavevector is $q=6.6$ in all curves. For both densities we show the directions $q_x=0$ and $q_y=0$ (solid lines, lying indistinguishable on top of each other) and the isotropic part of the $\beta$-correlator (dashed). Only for $\varepsilon=10^{-3}$, we also show the directions $q_x=q_y$ (upper dotted) as well as $q_x=-q_y$ (lower dotted). Inset: Focus on the anisotropy. Shown is the correlator as function of angle $\theta$ at time $t=10^5$, referenced to the one at $\theta=0$, for the two densities. The line (through the data) shows the result from the $\beta$-analysis for $\varepsilon=0$ which is proportional to $q_xq_y=q^2\cos(\theta)\sin(\theta)$.
}
\end{figure}

In Fig.~\ref{fig:fcbetaincoh}, we present the agreement between the full solutions for $\Phi_{\bf q}^s$ and the $\beta$-correlators near the critical plateau. The derived $\beta$-correlator is compared to $(\phis- f_q^{cs})/h^s_q$ for different directions of the wavevector ${\bf q}$. The positive slope of the correlation function for $q_x=q_y$ is hardly visible as the anisotropy in the $\beta$-process window predicted by theory is rather small. We conclude that shear flow frees the particle which would be localized in the quiescent glass initially in a rather isotropic process. 

Note that in Fig.~\ref{fig:fcbetaincoh}, the shape of the isotropic curves (solid lines) is independent of $|\bf q|$, since $\mathcal{G}^{(s,aniso)}_{\bf q} (t)=0$ there, giving rise to the well known factorization property. The shapes of the anisotropic curves (dotted lines) on the other hand do depend on  $|\bf q|$, i.e., the factorization does not hold. This statement can also be verified by Fig.~\ref{fig:vonq2}: The function $\mathcal{G}^{(s,aniso)}_{\bf q} (t)/\dot\gamma t$ does depend on $|\bf q|$. 

\subsection{$\alpha$-master-curves}\label{sec:alpha}
For $\varepsilon\geq0$ and $\dot\gamma\to 0$ with $\dot\gamma t=const.$, the correlators approach scaling functions $\Phi^{s+}_{\bf q}(\tilde t)$ (with $\tilde t=\sqrt{c^{(\dot\gamma)}/(\lambda-\frac{1}{2}})\dot\gamma t\equiv \tilde c \dot\gamma t$), which depend only on the timescale set by  $\dot\gamma$, i.e., they are independent of the short time dynamics set by $D_0$ \cite{Fuchs03}. The rescaled time $\tilde t$ actually corresponds to the accumulated strain since switch-on of shear, and the scaling law for $\Phi^{s+}_{\bf q}(\tilde t)$ expresses that the decorrelation is a function of the strain only. 
These functions obey a scaling equation, the so called $\alpha$ scaling equation. Its derivation (see App.~\ref{app:alpha}) is complicated by the fact that the memory function in Eq.~\eqref{eq:incofin} is not a function of $t-t'$, but of $t$ and $t'$ separately. Because of this, in the equation below, derivatives with respect to the advected wavevectors appear (with $\bar m_{\vct{q}}^{s}(t)$ defined in Eq.~\eqref{eq:mfin2}), 
\begin{align}
&\Phi_{\bf q}^{s+}(\tilde t)= \bar m_{\vct{q}}^{s+}(\tilde t)-\frac{d}{d\tilde t}\int_0^{\tilde t} d\tilde t' \bar m_{\vct{q}(\tilde t'/\tilde c)}^{s+}(\tilde t-\tilde t')\Phi_{\bf q}^{s+}(\tilde t')\nonumber\\
&+\int_0^{\tilde t} d\tilde t' \frac{\partial \vct{q}(\tilde t'/\tilde c)}{\partial \tilde t'}\cdot\left(\frac{\partial}{\partial \vct{q}(\tilde t'/\tilde c)}\bar m_{\vct{q}(\tilde t'/\tilde c)}^{s+}(\tilde t-\tilde t')\right) \Phi_{\bf q}^{s+}(\tilde t').\label{eq:alphasc}
\end{align}
The derivatives with respect to the advected wavevectors complicates also the numerical solution of this equation, but it shows that the correlator indeed obeys the scaling described above, also for the case when the memory function does not depend on $t-t'$ only. The reason is that the advected wavevectors causing the deviation from $t-t'$ naturally depend on the strain $\dot\gamma t$. It can be shown that the short time solution of Eq.~\eqref{eq:alphasc} at $\varepsilon=0$ is given by Eqs.~\eqref{eq:betafin} and \eqref{eq:eps=0}, 
\begin{equation}
\Phi_{\bf q}^{s+}(\tilde t\to 0)=f_q^{sc}-\tilde h^{s}_{\bf q} \tilde t,\label{eq:corrsh}
\end{equation}
with $\tilde h_{\bf q}^s=h^s_q(1+\alpha(|{\bf q}|) \frac{q_xq_y}{q^2} /\tilde c) $ (see Eq.~\eqref{eq:alphafunc}). 

\begin{figure}
\includegraphics[angle=270,width=0.95\linewidth]{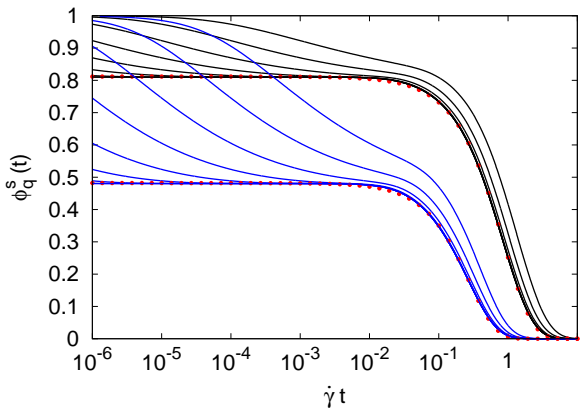}
\caption{\label{fig:incohglscal} Transient incoherent density correlator for $\varepsilon=10^{-3}$ and ${\bf q}d=6.6\, {\bf e}_y$ (upper curves) and ${\bf q}d=12.6\, {\bf e}_y$ (lower curves). Shear rates are $\dot\gamma=10^{-n}$ with $n=2, \dots, 9$ (for ${\bf q}d=6.6\, {\bf e}_y$, same data as in Fig.~\ref{fig:incohgl}). Here the time axis is scaled by shear rate to demonstrate the approach to the master function. Dots show fitted compressed exponentials with exponents $\mu=1.05$ (upper) and 1.13 (lower curve).}
\end{figure}

\begin{figure}
\includegraphics[width=0.9\linewidth]{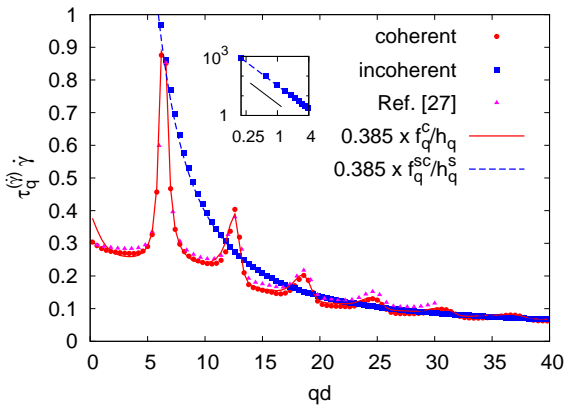}
\caption{\label{fig:tau} 
Relaxation time scale of the master-curve for $\varepsilon=10^{-3}$ and ${\bf q}=q {\bf e}_y$. The lines show the time scales as estimated from Eqs. \eqref{eq:taucoh} and \eqref{eq:tauincoh}. The inset shows the small $q$-data for the incoherent case in a logarithmic graph, demonstrating the divergence with $1/q^2$. The line shows the slope of -2.}
\end{figure}

The approach to the master function is exemplified in Fig.~\ref{fig:incohglscal}, where the correlators for a glassy state are plotted on a rescaled time axis.
We characterize the master functions by fitting to it compressed exponentials of the form
\begin{equation}
\lim_{\dot\gamma\to 0,\dot\gamma t=\mathcal{O}(1)}\phis=\Phi^{s+}_{\bf q}(\tilde t)\approx\tilde{f}_q\exp\left[-(t/\tau_{\bf q}^{(\dot\gamma)})^{\mu_{\bf q}}\right].\label{eq:fit}
\end{equation}
While the resulting value of the fit parameter $\tilde{f}_q$ is very close to $f_q$, this equality is not enforced by the fitting procedure.
Both the resulting relaxation timescale $\tau_{\bf q}^{(\dot\gamma)}$ as well as the stretching exponent $\mu_{\bf q}$ depend on the wavevector and the separation parameter $\varepsilon$. In Fig.~\ref{fig:tau}, we show the timescale for ${\bf q}$ pointing in $y$ direction as function of $|{\bf q}|$, for both coherent and incoherent correlators at $\varepsilon=10^{-3}$. The fit has been done with the data for $\dot\gamma=10^{-9}$. The coherent data are included in order to test and verify the good agreement to the data from Ref.~\cite{fabian_proc}, which were obtained on a Cartesian grid. The incoherent values of the time scale are as expected much smoother as a function of $q$, while for large $q$, the two cases approach each other.

This $q$ dependence of the timescale of the final decay is already visible in the $\beta$-correlator; Recalling its solution for ${\bf q}= q {\bf e}_y$ in Eq.~\eqref{eq:eps=0} and rewriting Eqs.~\eqref{eq:rec} and \eqref{eq:betafin} as the first order of an exponential decay from the plateau, $\Phi_q(t)\approx f_q^c\exp(-\tilde t h_q/f_q)$, we extract the time scale
\begin{align}
\tau_{q{\bf e}_y}^{(\dot\gamma)}=\frac{f^c_q}{|\dot\gamma|h_q}\sqrt{\frac{\lambda-\frac{1}{2}}{c^{(\dot\gamma)}}}\label{eq:taucoh}
\end{align} 
for the coherent, and
\begin{align}
\tau_{q{\bf e}_y}^{(\dot\gamma)}=\frac{f^{sc}_q}{|\dot\gamma|h^s_q}\sqrt{\frac{\lambda-\frac{1}{2}}{c^{(\dot\gamma)}}}\label{eq:tauincoh}
\end{align}
for the incoherent case.
These curves are also shown in Fig.~\ref{fig:tau}. We find that the forms \eqref{eq:taucoh} and \eqref{eq:tauincoh} indeed describe very well the $q$ dependence of the relaxation time scale. While the upper equations yield a prefactor of roughly $\sqrt{\frac{\lambda-\frac{1}{2}}{c^{(\dot\gamma)}}}=0.252$, we achieved the best agreement by setting it to 0.385. This difference is not unexpected since the relaxation time scale depends on $\varepsilon$, and we are comparing the values for $\varepsilon=0$ (Eqs. \eqref{eq:taucoh} and \eqref{eq:tauincoh}) to the one at $\varepsilon=10^{-3}$ (Fig.~\ref{fig:tau}).

\begin{figure}
\includegraphics[width=0.95\linewidth]{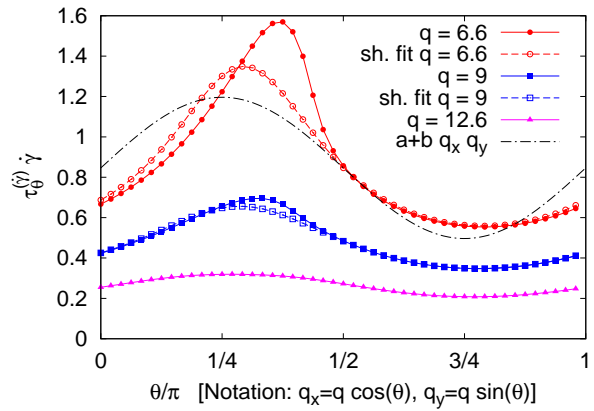}
\caption{\label{fig:tauanginco} 
Relaxation time scale of the incoherent master-curve as function of angle for $\varepsilon=10^{-3}$. Full symbols show the timescales for a fit of Eq.~\eqref{eq:fit} to the complete relaxation from the plateau to zero, i.e., including the regions where the functions show the shoulder-like deviations. Open symbols show the timescales obtained from fitting Eq.~\eqref{eq:fit} up to $\dot\gamma t=1$ (excluding the shoulders). These are not shown for $q=12.6$ since the two data sets are indistinguishable.}
\end{figure}

The relaxation timescale of the master-curves depends also on the direction of ${\bf q}$. This dependence is shown in Fig.~\ref{fig:tauanginco}, where $\tau_{\bf q}^{(\dot\gamma)}$ is plotted versus the angle $\theta$ (defined by $q_x=q\cos\theta$, $q_y=q\sin\theta$) for various values of $q$. We see that in most cases, a direction between $\theta=\pi/4$ and $\theta=\pi/2$ has the largest relaxation time. While the dependence on $q$ of the relaxation time scale can be well understood by the $\beta$ analysis (compare Fig.~\ref{fig:tau}), this is not quite true for the angular depedence: From the finding that $\mathcal{G}^{(s,aniso)}_{\bf q} (t)/\dot\gamma t$ in Eq.~\eqref{eq:betafin} is proportional to $q_xq_y$, we would expect that $\tau_{\bf q}^{(\dot\gamma)}\propto (a+b\sin \theta \cos\theta)$, where $a$ and $b$ describe the relative size of isotropic compared to anisotropic contributions. This functional form is also shown in Fig.~\ref{fig:tauanginco}. We see that the shape of $\tau_{\bf q}^{(\dot\gamma)}$ is quite different from this naive expectation, at least for small wavevectors, while the curve for the largest wave vector shown follows this simple form very well.
  
For small wavevectors, the correlators develop an angle-dependent shoulder at long times, and the shape of the curves is very different from a stretched exponential. These shoulders are an unexpected feature which is also seen in our simulations as shown in Sec.~\ref{sec:sim}. For the $\dot\gamma=10^{-9}$-curves used to create Fig.~\ref{fig:tauanginco}, these shoulders start to develop at roughly $t=10^9$. Fitting the curves up to $t=10^9$ ('short fit') yields the timescales shown as open symbols in Fig.~\ref{fig:tauanginco}. One sees that these are closer to the functional form $(a+b\sin \theta \cos\theta)$. Furthermore, since the difference between 'complete fit' and 'short fit' is a measure for the shoulder-like deviation from stretched exponentials, we note that the development of shoulders is most pronounced for small $q$ and for the direction near $\theta=3\pi/8$.
\begin{figure}
\includegraphics[angle=270,width=1\linewidth]{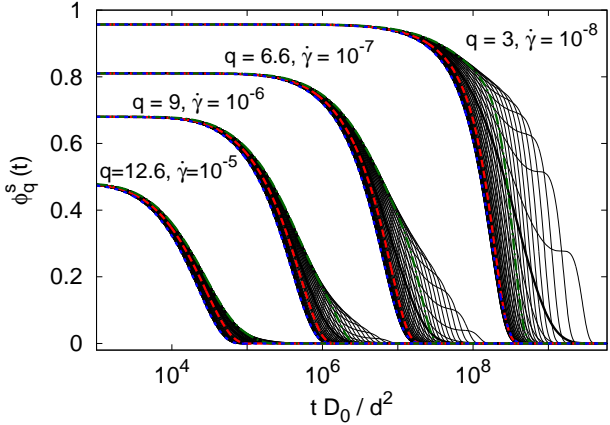}
\caption{\label{fig:allangles} 
Final decay for all angles of our numerical grid (48 curves for each wavevector). One can clearly see the shoulders for small $q$. We show different shear rates, as labeled, for visibility. The four directions of Fig.~\ref{fig:incohgl} are coded in the same way.}
\end{figure}
Following this discussion, we show in Fig.~\ref{fig:allangles} the final decay for all angles of our numerical grid. Shown are the three wavevectors from Fig.~\ref{fig:tauanginco}, and additionally $q=3$. For $q=3$ and $q=6.6$, the shoulders are best visible. They are present for a small range of angles (compare Fig.~\ref{fig:tauanginco}). We see that the height of the shoulders increases with decreasing wavevector. This can be explained by the fact that they appear for all $q$ at roughly the same strain ($\dot\gamma t\approx 1$) and the curves with large $q$ relax to zero before that time.

\section{Mean Squared Displacements}\label{sec:MSDs}
Knowing the equation for the incoherent density correlator under shear, we can now deduce from it the ones for the mean squared displacement (MSD) of the tagged particle for the different spatial directions and show their asymptotic solutions for long times. The transient MSDs so obtained describe a particle's motion after switching-on of shear at time $t=0$ averaged over equilibrium initial conditions. 

Before we start, we have to show the connection of the density correlator to the MSD, involving coordinates $a,b,c,d\in \{x_s,y_s,z_s\}$ of the particle at time $t$ or $t=0$. This MSD has to be formed with the conditional probability $W_2(\Gamma t,\Gamma'0)$, that the system is at state-point $\Gamma$ at time $t$ after it was at state-point $\Gamma'$ at $t=0$ \cite{Risken,Fuchs09}. The MSDs we will be looking for are of the form
\begin{multline}
 \left\langle[a(t)+\dot\gamma t b(t)-c(0)-\dot\gamma t d(0)]^2\right\rangle =\\\iint d\Gamma d\Gamma'\,
  [a(\Gamma)+\dot\gamma t b(\Gamma)-c(\Gamma')-\dot\gamma t d(\Gamma')]^2 W_2(\Gamma t,\Gamma'0) \,.
\end{multline}
It is a straight forward calculation to show that this mean squared displacement  is found by taking the limit of small $q$ of the corresponding correlator,
\begin{multline}
\left\langle[a(t)+\dot\gamma tb(t)-c(0)-\dot\gamma td(0)]^2\right\rangle\\=\lim_{q\to0}\frac{1-\left\langle e^{-iq (c+\dot\gamma td)}\ti e^{iq(a+\dot\gamma tb)} \right\rangle}{q^2}.\label{eq:defi}
\end{multline}
From this equation, we will be able to derive the desired MSDs. This will be done separately for the different directions, since the MSDs will be anisotropic, as was already seen in the low density case, Eq.~\eqref{eq:elrick}.
\subsection{Neutral Direction}
The calculation for the  neutral direction is in strong analogy to the equilibrium case \cite{Voigtmann04,Fuchs98}. Using Eq.~\eqref{eq:defi}, we see that we have to expand the correlator for ${\bf q}=q {\bf e}_z$ pointing in $z$-direction to get
\begin{equation}
\delta z^2(t)\equiv\left\langle[z(t)-z(0)]^2\right\rangle= 2\lim_{q\to0}\frac{1-\Phi_{q \vct{e}_z}^{s}(t)}{q^2} \label{eq:msdz}.
\end{equation}
$\delta z^2(t)$ is the transient mean squared displacement of the particle in $z$-direction. Its equation of motion is achieved by expanding \eqref{eq:incofin} to order $q^2$ and identifying the terms via \eqref{eq:msdz}. The equation is then integrated over time to get,
\begin{equation}
-\frac{1}{2} \delta z^2(t)+t-\frac{1}{2}\int_0^t dt'\int_0^{t'} dt''m_z^0(t'-t'')\partial_{t''}\delta z^2(t'')=0,
\end{equation}
with the memory function in the low $q$ limit (see Eqs.~\eqref{eq:mfMSD} and \eqref{eq:mfMSD2} for the definition of $F(\vct{k},\vct{q},t)$)
\begin{equation}
m_z^0(t)=\sum_{{\bf k}}k_zk_z\, F(\vct{k},\vct{0},t).\label{eq:memz}
\end{equation}
Since $m_z^0(t)$ has only one time-argument, one can rewrite the above equation using the standard trick of partial integrations and $\delta z^2(t=0)=0$,
\begin{equation}
\delta z^2(t)+\int_0^t dt'm_z^0(t-t')\delta z^2(t')=2t.\label{eq:asineq}
\end{equation}
Eq.~\eqref{eq:asineq} now looks similar to the equilibrium case \cite{Fuchs98}, and its schematic version has been studied before \cite{Krueger09d,Zausch08,Krueger10b}.
The long time limit of the solution corresponds to the small-$z$ part of its Laplace transform $\delta z^2(z)=\int_0^\infty dt\, e^{-z\,t} \delta z^2(t)$. The convolution theorem can be applied. We find for $t\to\infty$,
\begin{equation}
\lim_{t\to\infty} \delta z^2(t)= \frac{2t}{1+m_z^0(z=0)}.\label{eq:zlong}
\end{equation}
In contrast to the equilibrium case, $m_z^0(z=0)$ is always finite under shear and the MSD is always diffusive at long times. 
In the glass, we have $\lim_{\dot\gamma\to0}m_z^0(z=0)\propto|\dot\gamma|^{-1}$ (compare Eqs.~(\ref{eq:taucoh},\ref{eq:tauincoh}) and the $\alpha$-scaling equation in Ref.~\cite{Fuchs03}) leading to the scaling relation at small shear rates,
\begin{equation}
\lim_{t\to\infty} \delta z^2(t)= 2 \beta_z |\dot\gamma| t,\label{eq:scalingy}
\end{equation}
where the coefficient $\beta_z=(|\dot\gamma| m_z^0(z=0))^{-1}$ is asymptotically independent of shear rate as $\dot\gamma\to0$. We see that the long time diffusivity $D_z^{(\dot\gamma)}=\beta_z|\dot\gamma|$ is then proportional to the shear rate and independent of the short time diffusivity $D_0$. Shear flow thus enables the particle to diffuse also perpendicular to the flow, which highlights that flow melts the glass. The affine average particle motion decorrelates the non-ergodic structural relaxation. It becomes ergodic in all directions and for all variables that would be non-ergodic in the glass.

The same linear scaling of the diffusion coefficient with $\dot\gamma/d^2$ is also predicted for sheared non-Brownian particles \cite{Sierou04}, yet the range of shear rates for these predictions is very different. The above analysis holds for Pe$_0\ll1$, while the limit of non-Brownian particles is approached for Pe$_0\gg1$ \cite{Sierou04}. Presumably, also the physical mechanisms differ. For Pe$_0\ll1$, shear destroys the localization of particles in a quiescent glass and causes structural relaxation. The relevant length scale is the localization length that can be read off from the quiescent MSD and corresponds to the Lindemann length at solidification; often it is connected to the picture of 'cages'. For Pe$_0\gg1$ shear dominates over Brownian motion on all length scales except for in a narrow boundary layer close to particle contact. 
\subsection{Gradient Direction}
The derivation for the gradient direction is similar to the neutral direction. The correlator with $\bf q$ pointing in $y$-direction, ${\bf q}=q {\bf e}_y$, is expanded, 
\begin{equation}
\delta y^2(t)\equiv\left\langle[y(t)-y(0)]^2\right\rangle= 2\lim_{q\to0}\frac{1-\Phi_{q \vct{e}_y}^{s}(t)}{q^2} \label{eq:msdy}.
\end{equation}
The equation of motion follows analogously and reads
\begin{equation}
\delta y^2(t)+\int_0^t dt'm_y^0(t-t')\delta y^2(t')=2t,\label{eq:ymsdeq}
\end{equation}
with the memory function  
 \begin{equation}
m_y^0(t)=\sum_{{\bf k}}k_y(t)k_y\, F(\vct{k},\vct{0},t).
\end{equation}
Note the slight difference in this memory function compared to the one in Eq.~\eqref{eq:memz}: One of the $k_y$ is time dependent.
As expected, the long time limit of $\delta y^2(t)$ is given by 
\begin{equation}
\lim_{t\to\infty} \delta y^2(t)= \frac{2t}{1+m_y^0(z=0)}.\label{eq:ydir}
\end{equation}
This leads to a scaling relation similar to Eq.~\eqref{eq:scalingy} for glassy states at low shear rates, 
\begin{equation}                                                                        \lim_{t\to\infty} \delta y^2(t)= 2 \beta_y |\dot\gamma| t\equiv 2D_y^{(\dot\gamma)}t.\label{eq:alphadef}
\end{equation}
We have no reason to expect that the coefficients $\beta_y=(|\dot\gamma| m_y^0(z=0))^{-1}$ and $\beta_z$ are equal, i.e., $D_y^{(\dot\gamma)}=\beta_y|\dot\gamma|$ will take a different number compared to $D_z^{(\dot\gamma)}$. Indeed, these have been found slightly different in simulations \cite{Varnik08,Zausch08}. Otherwise, the qualitative discussion of the physical mechanism behind Eq.~\eqref{eq:alphadef} can be taken over from the neutral direction.
\subsection{Flow Direction -- Glass Taylor Dispersion}\label{sec:flow}
Concerning the MSD in flow direction, we have to note that we are seeking space-translational invariant quantities. The expression $\left\langle [x(t)-x(0)]^2\right\rangle$ is not translationally invariant and hence not appropriate (it depends on $y(0)$, see Eq.~\eqref{eq:elrick}). Quantities which fulfill this invariance are $\left\langle [x(t)-\dot\gamma t y(t)-x(0)]^2\right\rangle$ and $\left\langle [x(t)+\dot\gamma t y(0)-x(0)]^2\right\rangle$. One can show that the two are identical for small densities
\begin{multline}
\left\langle [x(t)-\dot\gamma t y(t)-x(0)]^2\right\rangle=\left\langle [x(t)+\dot\gamma t y(0)-x(0)]^2\right\rangle\\=2D_0t+\frac{2}{3}D_0\dot\gamma^2 t^3.\label{eq:lowd}
\end{multline}
Comparing to Eq.~\eqref{eq:elrick}, we see that the drift term $y(0)^2\dot\gamma^2 t^2$ stemming from constant motion with velocity $y(0)\dot\gamma t$ is absent. It depends on $y(0)$ and has to be missing in our translationally invariant formulation.

For finite densities, we can not expect the two definitions in the first line of Eq.~\eqref{eq:lowd} to still be identical, their difference stays in fact unknown. Our approach naturally leads to defining the MSD for the $x$-direction in terms of our transient density correlator,
\begin{align}
\delta x^2(t)&\equiv \left\langle [x(t)-\dot\gamma t y(t)-x(0)]^2\right\rangle\notag\\
&=2\lim_{q\to0}\frac{1-\Phi_{q \vct{e}_x}^{s}(t)}{q^2},\label{eq:msdx}
\end{align}
with
\begin{equation}
\Phi_{q \vct{e}_x}^{s}(t)=\left\langle e^{-iqx_s}\ti e^{iqx_s}e^{-\dot\gamma t iqy_s}\right\rangle.
\end{equation}
This definition agrees with the formal one in Eq.~\eqref{eq:defi}.
The equation for $\delta x^2(t)$ can now be gained by expanding the equation for the correlator $\Phi_{q \vct{e}_x}^{s}(t)$ in $q$, 
\begin{align}
\partial_t\delta x^2(t)+\int_0^{t}dt' m_x^0(t,t')
\partial_{t'} \delta x^2(t')=2\frac{\Gamma^s_{q\vct{e}_x}(t)}{q^2}.\label{eq:xricht}
\end{align}
with
\begin{equation}
m_x^0(t,t')=\sum_{\vct{k}}\left[k_x-\dot\gamma tk_y(t-t')\right] \frac{k_x-\dot\gamma t' k_y}{1+(\dot\gamma t')^2} F(\vct{k},\vct{0},t-t').\label{eq:mxricht}
\end{equation}
Because of (compare Eq.\eqref{eq:spec}) 
\begin{equation}
2\frac{\Gamma^s_{q\vct{e}_x}(t)}{q^2}=2+2\dot\gamma^2 t^2,\label{eq:init}
\end{equation}
we recover the low density limit of Eq.~\eqref{eq:lowd} using $m_x^0(t,t')\equiv0$, as required for non-interacting particles (infinite dilution). 
Because the memory function in Eq.~\eqref{eq:xricht} is not a function of the difference of its arguments only, the analysis of the leading long time terms of $\delta x^2$ for dense systems involves a bit more work compared to the other directions above, see App.~\ref{app:taylor}. We find
\begin{equation}
\lim\limits_{t\to\infty}\delta x^2(t)=\frac{2\dot\gamma^2}{3+3m_y^0(z=0)} t^3.\label{eq:Taylor1}
\end{equation} 
This result deserves some discussion: It can be regarded as the Taylor dispersion for Brownian particles in a shear melted glass. The MSD in $x$-direction grows cubically in time as it does for small densities. The intriguing result is that the coefficient for the $t^3$ term is connected to the long time diffusivity for the  $y$-direction in the same way as in the low density limit. This can be further illustrated by writing
\begin{equation}
\lim\limits_{t\to\infty}\delta x^2(t)=\frac{2}{3}\delta y^2(t)\dot\gamma^2 t^2=\frac{2}{3}D_y^{(\dot\gamma)} \dot\gamma^2 t^3,\label{eq:Taylorgl}
\end{equation}
which holds identically in the low density limit, Eq.~\eqref{eq:elrick}, and was also found in Ref.~\cite{Sierou04} for non-Brownian particles. We see that this relation comes about because for long times, $\delta x^2$ is governed by $m_y^0(t-t')$, see Eq.~\eqref{eq:longtime}. This is physically plausible if we recall the reason for the $t^3$-term: If the particle moves in $y$-direction, it gets a ``boost'' in $x$-direction due to the shear flow. It is hence not surprising that the $t^3$ term is proportional to $D_y^{(\dot\gamma)}$, but the result that the very same relation holds as in the low density limit is nontrivial and unexpected. 

Despite the similarities of the glass Taylor dispersion and the low density one, there is an important difference: In glasses, the long time term in Eq.~\eqref{eq:Taylorgl} is independent of the bare diffusivity $D_0$ (set to unity here) and obeys the yield scaling law, 
\begin{equation}
\lim\limits_{t\to\infty}\delta x^2(t)=\frac{2}{3}\beta_y\dot\gamma^2 |\dot\gamma| t^3,\label{eq:scalingx}
\end{equation}
again, with the same $\beta_y$ as in Eq.~\eqref{eq:alphadef}.  It is also possible to derive the next order term in $\delta x^2(t)$, see again App.~\ref{app:taylor}. It reads 
\begin{equation}
\frac{\dot\gamma^2}{1+m_y^0(z=0)}\frac{\left[\partial_z m_y^0(z=0)\right]+\frac{m_{xy}^0(z=0)}{\dot\gamma}}{1+m_y^0(z=0)}t^2.\label{eq:bt}
\end{equation}
Such a term proportional to $t^2$ is not present in the low density limit, Eq.~\eqref{eq:lowd}. It comes about because the memory function is not a function of $t-t'$. Recall that we are currently calculating the {\it transient} MSD. The stationary MSD might not have a term of order $t^2$ for $t\to\infty$. Note that the term in Eq.~\eqref{eq:bt} is $\propto \dot\gamma^2 t^2$ for glassy states. 
\subsection{Cross Correlation}
In the system under shear, there is a correlation between $x$ and $y$ which is not present without shear, see Eq.~\eqref{eq:elrick}. In our translationally invariant formulation, we define it the following way
\begin{equation}
\delta xy(t)\equiv \left\langle \left[x(t)-x(0)-\dot\gamma t y(t)\right]\left[y(t)-y(0)\right]\right\rangle.\label{eq:xydef}
\end{equation}
It can be derived considering the correlator for the diagonal direction 
${\bf q}(t=0)=\left(q , q , 0\right)^T$
leading to 
\begin{eqnarray}
\delta xy(t)&=&\lim_{q\to0}\frac{1-\Phi_{q(\vct{e}_x+\vct{e}_y)}^{s}(t)}{q^2} - \frac{\delta x^2(t)+\delta y^2(t)}{2}.
\label{eq:xy2}
\end{eqnarray}
See App.~\ref{app:cross} for the derivation of the long time result of Eq.~\eqref{eq:xy2}.  The leading order of $\delta xy(t)$ is proportional to $t^2$ as in the low density case,
\begin{equation}
\lim_{t\to\infty}\delta xy(t)=-\frac{\dot\gamma}{1+m_y^0(z=0)}t^2=-D_y^{(\dot\gamma)}t \dot\gamma t.\label{eq:xyres}
\end{equation}
The last step followed with the result for the long time diffusion in $y$-direction in Eq.~\eqref{eq:ydir}. 
We see that $\delta y^2(t)$ and $\delta x y(t)$ are related to each other as in the low density limit, except for the minus sign. This sign originates from our definition in Eq.~\eqref{eq:xydef}. Note that defining $\delta xy(t)=\left\langle \left[x(t)-x(0)+\dot\gamma t y(0)\right]\left[y(t)-y(0)\right]\right\rangle$ instead would yield a plus sign in \eqref{eq:xyres}. The simulations described below also give this sign difference depending on definition.

The scaling relation in glassy states as $\dot\gamma\to0$ follows,
\begin{equation}
\lim_{t\to\infty}\delta xy(t)=-\beta_y |\dot\gamma| \dot\gamma t^2+\mathcal{O}(t),\label{eq:cross}
\end{equation}
with $\beta_y$ as in Eq.~\eqref{eq:alphadef}. The sign of $\delta xy(t)$ depends on the sign of $\dot\gamma$, which is expected since inverting the direction of shearing corresponds to inverting either $x$ or $y$.
\section{Numerical Results for the Mean Squared Displacements}\label{sec:MSDnum}
After having solved the equations for the incoherent correlator $\phis$, we can solve numerically Eqs.~\eqref{eq:ymsdeq}, \eqref{eq:xricht} and \eqref{eq:deltaxy}, for the MSD in $y$ and $x$ directions as well as the cross correlation. In the 2D numerical calculation we can of course not discuss the MSD for the neutral direction.  

\begin{figure}
\includegraphics[angle=270,width=0.9\linewidth]{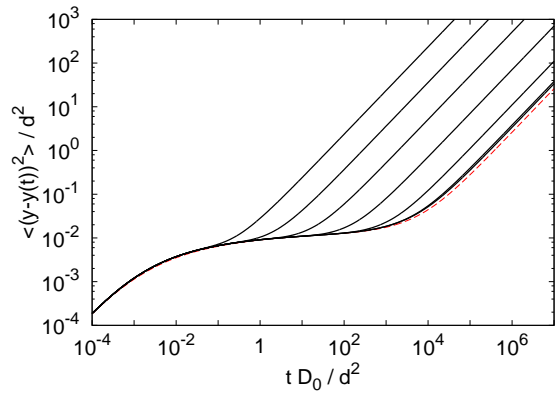}
\caption{\label{fig:msdfl}
Mean squared displacement for the gradient direction for different values of the shear rate, $\dot\gamma=10^{-n}$ with $n=0, \dots, 6$ and   $\varepsilon=-10^{-3}$. The curve for $\dot\gamma=10^{-6}$ cannot be distinguished from the equilibrium curve. The dashed curve shows the equilibrium MSD from Ref.~\cite{Bayer07}.}
\end{figure}

In Fig.~\ref{fig:msdfl}, we show the MSD for the gradient direction for different shear rates in a fluid state ($\varepsilon<0$). As was  discussed in Sec.~\ref{sec:corr}, the MSD approaches the equilibrium MSD for $\dot\gamma\to0$, the curve for $\dot\gamma=10^{-6}$ cannot be distinguished from it. In Fig.~\ref{fig:msdfl}, we also show the equilibrium MSD for the same $\varepsilon$ taken from Ref.~\cite{Bayer07}. The slight disagreement at long times is due to the different grids chosen, as discussed above.

\begin{figure}
\includegraphics[angle=270,width=0.9\linewidth]{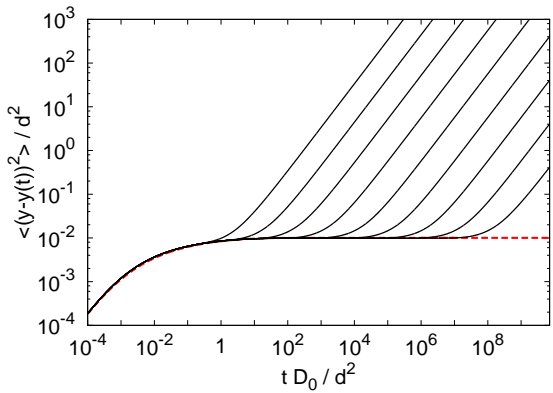}
\caption{\label{fig:msd}
Mean squared displacement for the gradient direction for different values of the shear rate, $\dot\gamma=10^{-n}$ with $n=1, \dots, 9$ and   $\varepsilon=10^{-3}$. The dashed curve shows the equilibrium MSD from Ref.~\cite{Bayer07}.}
\end{figure}

For the glassy state ($\varepsilon\ge0$), the long time diffusivity $D_y^{(\dot\gamma)}$ (defined below Eq.~\eqref{eq:alphadef}) is governed by shear for arbitrarily small shear rates. In the limit of $\dot\gamma\to0$, the scaling law of Eq.~\eqref{eq:alphadef} is approached with $\beta_y$ approaching a constant. Glass curves for $\varepsilon=10^{-3}$ are shown in Fig.~\ref{fig:msd}.

\begin{figure}
\includegraphics[angle=270,width=0.9\linewidth]{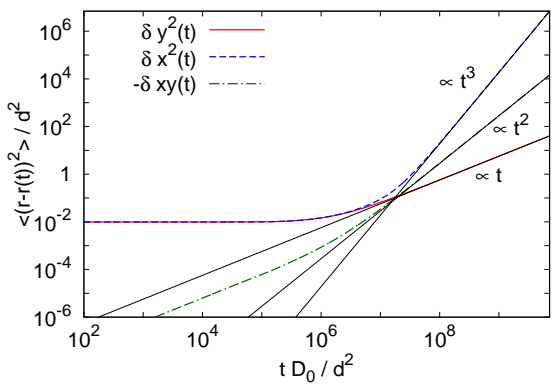}
\caption{\label{fig:msdani}
Glass Taylor dispersion: The MSDs for different directions for a glassy state, $\varepsilon=10^{-3}$ and shear rate, $\dot\gamma=10^{-7}$. Thin lines show the long times asymptotes, according to Eqs.~\eqref{eq:alphadef}, \eqref{eq:scalingx} and \eqref{eq:cross}.}
\end{figure}

In Fig.~\ref{fig:msdani} we finally compare the different directions and demonstrate the glass Taylor dispersion Eq.~\eqref{eq:Taylorgl}.  We see that the MSD for the $x$-direction cannot be distinguished from the one for the $y$-direction as long as $\dot\gamma t\ll 1$. For long times, with $\dot\gamma t=\mathcal{O}(1)$, the two functions separate and the one for the $x$-directions approaches the long time $t^3$ asymptote from Eq.~\eqref{eq:Taylorgl}.

In Fig.~\ref{fig:msdani}, also the cross correlation $-\delta xy(t)$ is shown. It is small compared to $\delta x^2$ and $\delta y^2$ for $\dot\gamma t\ll1 $ and approaches the asymptotic law Eq.~\eqref{eq:cross} for $\dot\gamma t=\mathcal{O}(1)$.

\section{Stationary Mean Squared Displacements}\label{sec:MSDstat}
In the previous sections, we derived the equations of motion for the transient MSDs. It has been noticed that these differ from the stationary ones \cite{Zausch08}. 
Before we discuss these differences, let us emphasize the similarities between transient and stationary MSDs giving rise to the lowest order approximation of setting them equal \cite{Fuchs09}. For long times, when the transient MSD has reached its linear (steady) dependence on time, it has to follow the long time diffusivity of the steady state (the system is then obviously in the steady state). In this regime, transient and stationary MSDs must hence approach each other. Consequently, long time  diffusivities as extracted from transient or stationary MSDs must be identical. The ITT approach of deriving the transient quantities thus proves very useful here: The results in Eqs.~\eqref{eq:zlong}, \eqref{eq:ydir}, \eqref{eq:Taylorgl} and \eqref{eq:xyres} hold for the stationary MSDs as well. 

\begin{figure}
\includegraphics[angle=270,width=0.9\linewidth]{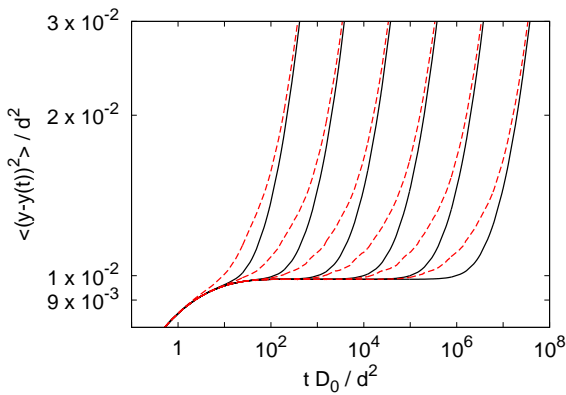}
\caption{\label{fig:msdstat}
Comparison of stationary (dashed) and transient (solid) MSDs for the gradient direction for a glassy state, $\varepsilon=10^{-3}$, and shear rate, $\dot\gamma=10^{-n}$ with $n=3,\dots, 8$. The necessity to take derivatives in Eq.~(\ref{eq:statMSD}) introduces small wiggles.}
\end{figure}

 In Ref.~\cite{Krueger10b}, an approximate relation between stationary and transient MSDs was derived, which builds on the waiting time derivative introduced in Ref.~\cite{Krueger09}. For directions perpendicular to the direction of shear, we found for the stationary MSD  $\delta z_s^2(t)$ (or $\delta y_s^2(t)$) in terms of the transient one introduced in Eqs.~\eqref{eq:msdz} and \eqref{eq:msdy},
\begin{multline}
  \delta z^2_{s}(t)\approx \delta z^2(t)
  +\tilde\sigma\frac d{dt}
  \left(\delta z^2(t)-\delta z^2_e(t)\right)\,.
  \label{eq:statMSD}
\end{multline}
$\delta z^2_e(t)$ denotes the MSD of the quiescent system without shear. The pre-factor $\tilde\sigma$  is the normalized integrated shear modulus,
\begin{equation}\label{eq:shmod}
  \tilde\sigma=\int_0^{\infty}\frac{\langle\sigma_{xy}e^{\Omega^\dagger s}
  \sigma_{xy}\rangle}{\langle\sigma_{xy}\sigma_{xy}\rangle}\,ds\,,
\end{equation}
with shear stress  $\sigma_{xy}=-\sum_iF_i^x y_i$.
For the case of hard spheres, the integration in \eqref{eq:shmod} has to be renormalized since the initial value $\langle\sigma_{xy}\sigma_{xy}\rangle$ diverges \cite{Fuchs05,Krueger10b}. During this renormalization, a free parameter $\kappa$ enters the equation for $\tilde\sigma$, which is independent of shear rate and density. The final expression for the stationary MSD for hard spheres is hence given by
\begin{align}
\tilde\sigma &\approx \kappa \int_0^{\infty}\langle\sigma_{xy}e^{\Omega^\dagger s}
  \sigma_{xy}\rangle\notag\\&\approx \frac{\kappa}{2}\int_0^\infty ds\int \frac{d^3 k}{(2\pi)^3} \frac{k_x^2 k_y(-s)k_y}{kk(-s)}\frac{S_k' S_{k(-s)}'}{S^2_k}\Phi^2_{\vct{k}(-s)}(s). \label{eq:modmct}
\end{align}
The last line followed with the ITT expression for the stationary stress, see Ref.~\cite{Fuchs09}. We can now evaluate Eq.~\eqref{eq:statMSD} with use of Eq.~\eqref{eq:modmct}. We used $\kappa=9\times10^{-3}$, as estimated from comparing with Ref.~\cite{Krueger10b}. Since Eq.~\eqref{eq:statMSD} holds only for directions perpendicular to the shear direction, we can only apply it to the $y$-direction in our 2-dimensional numerical analysis. This is shown in Fig.~\ref{fig:msdstat} for a glassy state at different shear rates. For the five shear rates given, we have $\dot\gamma\tilde\sigma=0.038 \pm 0.001$ (increasing with shear rate). These values compare well to the value of $\dot\gamma\tilde\sigma=0.04$ used to fit the Brownian dynamics simulation data in Ref.~\cite{Krueger10b}. We see, that the difference between transient and stationary correlators is most pronounced at intermediate times, whereas for short and long times, the two functions coincide.

\section{Comparison to simulations}\label{sec:sim}
In this section, we will  compare our theoretical MCT-ITT results to our simulations. Since MCT-ITT is a quantitative theory, there is no fit parameter to be adjusted. As will be illustrated, the simulation results show many unexpected features, some of which are captured qualitatively, but not quantitatively by MCT-ITT. The quantitative disagreement can be mostly explained by the underestimation of the overshoot scenario of the stress after switch on \cite{Krueger10b,Zausch08}, as will be discussed in detail.

The simulation algorithm is an event driven algorithm which describes the dynamics of hard particles, i.e. hard discs in the 2D case considered. It has been described in detail for 3D in Ref.~\cite{scala02} and its adaptation to two dimensions can be found in Ref.~\cite{fabian_proc}.
We consider a binary mixture of hard discs with the diameters of $d_l \equiv d$ and $d_b=1.4 d$  with equal particle number concentrations and a total amount of $N=N_l+N_b=1000$ hard discs in a 2D simulation box of volume $V$. Thus the packing fraction is given by $\eta = \pi \frac{N}{8V} (d_b^2 + d_l^2)$.
For this system, we find the glass transition point to be at roughly $\eta^c \approx 0.7948$ \cite{Weysser11}. 

Simulations have been performed at packing fractions of $\eta = 0.79$ (liquid) and $\eta =0.81$ (glass), discretizing the time in steps of $\tau_B = 5 \times 10^{-5}$ in units of $d^2/D_0$.

In the liquid ($\eta =0.79$), for the transient correlation functions, $300$ independent initial configurations were prepared and equilibrated for a time $\tau=10^{5}$ (using, only for equilibration, a Newtonian dynamics algorithm) which is large compared to the $\alpha$-relaxation time of $\tau_\alpha \approx 10^{3}$, given in units of $d^2/D_0$. For the stationary correlation functions $600-800$ independent initial configurations (depending on the shear rate) were prepared. According to the findings in Ref.~\cite{Krueger10b}, stationarity was assumed after $\dot \gamma t_w > 1$, where $t_w$ is the waiting time after switch on of shear.

Above the glass transition density, the preparation of transient correlators is nontrivial because the system without shear equilibrates very slowly. So for the glassy systems ($\eta =0.81$) we prepared $150$ independent equilibrated sets for the transient correlation functions.  Equilibration was achieved by waiting for a period $\tau=1.6\times10^6$, corresponding to an average displacement of the particles of half their diameters. After this waiting time, the correlation functions are independent of waiting time. We estimated the $\alpha$-relaxation time to be roughly $20 \%$ of our equilibration time. This is large compared to the time window examined in the following, so that this density can be regarded glassy in our simulation time window.

For the glassy stationary correlation functions, $150-300$ independent initial configurations (depending on shear rate) were prepared. Again stationarity was assumed after $\dot \gamma t_w> 1$.
\subsection{Correlators}\label{sec:Corr}

Let ${\bm r}_{i_j}(t_k)$ be the position of the $i$-th particle in the $j$-th of a total $M$ simulation sets for a given time $t_k$. Then the correlators at time $t_k$ for waiting time $t_w$ are calculated via
\begin{align}
\notag\Phi_{\bf q}^s(t_k,t_w) =  \frac{1}{M} \sum \limits_{j=1}^{M} \Biggl( \frac{1}{N} \sum \limits_{i=1}^N \exp \biggl[ i \bigl( {\bm q } \cdot{\bm r}_{i_j}(t_k+t_w) \\- q_x \dot \gamma t_k  y_{i_j}   (t_k+t_w) - {\bm q} \cdot{\bm r}_{i_j}(t_w) \bigr) \biggr] \Biggr)\label{eq:corrsim},
\end{align}
where external shear is switched on at the time origin, so that $t_w=0$ corresponds to the transient correlator. 

\begin{figure}
\includegraphics[clip = true, width=0.9\linewidth]{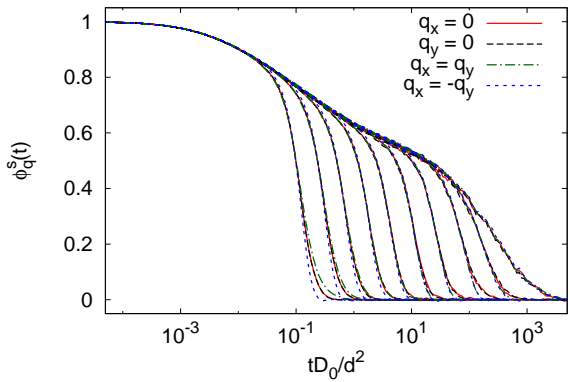}
\caption{Transient incoherent density correlators for $\eta= 0.79$ (liquid) and $|{\bm q}|d=6.5$ for different directions as labeled. Pe$_0$ numbers are $2 \times 10^0, 6 \times 10^{-1},  2 \times 10^{-1},  6 \times 10^{-2}, 2 \times 10^{-2}, 6  \times 10^{-3}$, $2 \times 10^{-3},  6 \times 10^{-4} ,2 \times 10^{-4}$ and Pe$_0=0$ from left to right.}  \label{corr-simu-dir-liquid}
\end{figure}

In Fig.~\ref{corr-simu-dir-liquid}, we show the transient correlation function for the liquid ($\eta = 0.79$) at different shear rates. We see the close analogy to the theoretical curves in Fig.~\ref{fig:incohfl}. For large dressed Peclet numbers Pe$=\dot\gamma\tau_\alpha$, the final decay is dominated by shear and the curves are anisotropic. As in Fig.~\ref{fig:incohfl}, the direction $q_x=0$ is (slightly) slower than the $q_y=0$ direction which is slower than the $q_x=-q_y$ direction for all shear rates. The correlator in $q_x=q_y$ direction shows a strong shear rate dependent behavior in its relaxation time: It decays as fast as the one for the $q_x=-q_y$ direction for small Pe$_0$ numbers, but exhibits the slowest relaxation time for large Pe$_0$ numbers. The plateau in Fig.~\ref{corr-simu-dir-liquid} is lower compared to Fig.~\ref{fig:incohfl} which we attribute to the bidispersity of the simulations. MCT calculations for binary hard discs in two dimensions for the quiescent system yield a plateau value of $f_q^{s,c} \approx 0.61 $ for $qd=6.5$ while the simulation yields $f_q^{s,c} \approx 0.58 $. \cite{hajnal-private}.

\begin{figure} \includegraphics[clip = true, width=0.9\linewidth]{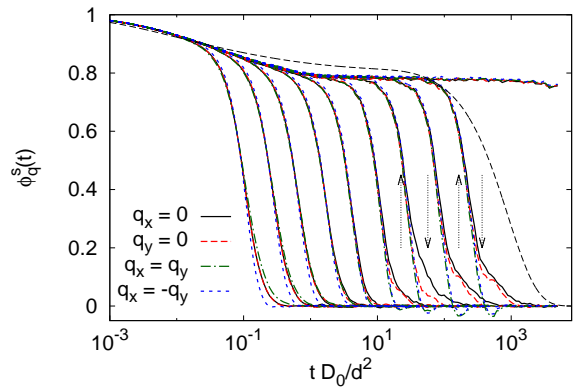}
\caption{Transient incoherent density correlators for $\eta= 0.81$ (glass) and $|{\bm q}|d=6.5$ for different directions as labeled. Pe$_0$ numbers are $2 \times 10^0, 6 \times 10^{-1},  2 \times 10^{-1},  6 \times 10^{-2}, 2 \times 10^{-2}, 6  \times 10^{-3}$, $2 \times 10^{-3},  6 \times 10^{-4} ,2 \times 10^{-4}$  and Pe$_0=0$ from left to right. For Pe$_0=2 \times 10^{-3}$ and $2 \times 10^{-4}$, arrows mark the stress-maximum (pointing up) and the time when the stress has reached its final value (pointing down); from Ref.~\cite{weyssertbp}. Dashed line is the MCT-ITT result for $q=6.6$, $q_x=0$, $\varepsilon=10^{-3}$ and Pe$_0=10^{-3}$. \label{corr-simu-dir-glass}}
\end{figure}

Fig.~\ref{corr-simu-dir-glass} shows the same functions for the glassy density ($\eta = 0.81$). These curves are in analogy to Fig.~\ref{fig:incohgl}. Additionally to the discussion of the liquid curves, we observe the emergence of shoulders for the smallest shear rates: For the directions  $q_x=0$ and $q_y=0$, the correlators drastically slow down at the end of the final relaxation process. We attribute this slowing down to the slowing down of the system after the stress overshoot, see below, and Refs.~\cite{Zausch08,Krueger10b} for the discussion of the shear stress overshoot scenario.

 Remarkably, we observe these shoulders  in MCT-ITT, compare Figs.~\ref{fig:tauanginco} and \ref{fig:allangles}. Figs.~\ref{fig:tauanginco} and \ref{fig:allangles} also show that MCT-ITT predicts them  most pronounced at a direction between $q_x=q_y$ and $q_y=0$,  this is why they are not as clearly seen in Fig.~\ref{fig:incohgl}.

Figure \ref{corr-simu-dir-glass} also presents one MCT-ITT curve for roughly the same parameters as the slowest of the simulation curves for a quantitative comparison. The small difference in plateau heights is as expected (we are comparing simulations for a binary mixture to theory for a mono-disperse system). Apart from that, the time scale of the initial deviation from the glassy plateau for the Pe$_0=10^{-3}$ curve agrees well with that of the Pe$_0 = 2 \times 10^{-4}$ simulation curves, i.e., MCT-ITT differs at most by a factor of five in shear rate. But for larger times, the simulation curves are much steeper (compare the compressing exponents in Fig.~\ref{fig:betasim} below) compared to the theory. We attribute this effect of large compressing exponents to the stress overshoot scenario after switch-on. While the theory curves qualitatively capture this compressing effect (the exponents in Fig.~\ref{fig:betasim} are greater than unity), it quantitatively underestimates it. E.g., the memory function in Eq.~\eqref{eq:incofin} does become negative for certain parameters (leading to slightly negative correlators at long times in Figs.~\ref{fig:incohgl} and \ref{fig:incohfl}), but the effect is much smaller compared to the simulation. As is seen in Fig.~\ref{corr-simu-dir-glass}, the described underestimation leads to a larger deviation of the curves at long times (compare also Fig.~\ref{fig:tausim} below).  

While the stress after switch on for the glassy state  will be presented elsewhere \cite{weyssertbp}, in order to underpin the conclusions of this section, we marked two characteristic times in Fig.~\ref{corr-simu-dir-glass}, namely the stress (overshoot) maximum as well as the time where it has approached its (lower) final value. First, we see that the MCT-ITT curve indeed starts deviating from the simulation curves at roughly the time, where the stress is at its maximum, underestimating the successive fast decay. Second, the shoulders indeed start emerging when the stress has reached its final value, and the dynamics seems to slow down drastically.

\subsection{$\beta$-regime}\label{sec:simbeta}
\begin{figure}
\includegraphics[clip = true, width=0.9\linewidth]{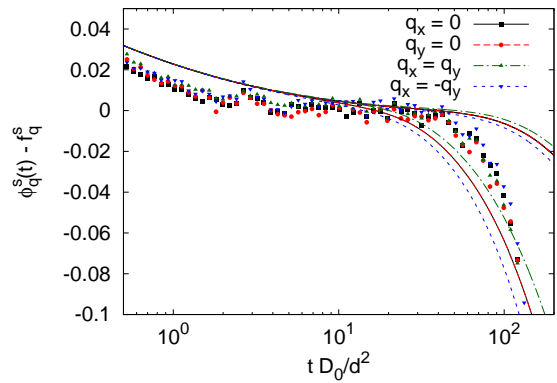}
\caption{The incoherent transient correlator near the plateau for $\eta=0.81$, Pe$_0=2 \times 10^{-4}$ and $|{\bm q}|d=6.5$ (simulations, datapoints). Lines show theory curves, with $\varepsilon=10^{-3}$, $|{\bm q}|d=6.6$, and Pe$_0=2 \times 10^{-4}$ (right set) and Pe$_0= 10^{-3}$ (left set). \label{fig:betafsim}}
\end{figure}

Following the discussion in Sect.~\ref{sec:beta}, it is interesting to compare simulation and theory in the $\beta$-regime, where the correlators are near the glassy plateau. For simplicity, we consider only a single wavevector, focussing on quantitative comparison rather than testing factorization properties. For this test, it is sufficient to regard  $\Phi_{\bf q}^s(t)-f_q$, where the plateau value $f_q$ was chosen appropriately in simulation and theory. Dividing by $h_q^s$ (compare Fig.~\ref{fig:fcbetaincoh}) is only necessary when testing factorization properties or comparing to the $\beta$ correlator, here it would only lead to a stretching of the $y$-axis. Fig.~\ref{fig:betafsim} shows the curves for $q=6.6$, restating that the time scale for the initial decay from the plateau is quantitatively described by MCT-ITT with a multiplicative error between 1 and 5. Additionally, we observe that the initial anisotropy predicted by MCT-ITT ($\propto q_xq_y$, compare Fig.~\ref{fig:fcbetaincoh}) is within the statistical noise of the simulations and a detailed comparison has to be left for future work.

\subsection{$\alpha$ Master Curves}\label{sec:simalpha}
\begin{figure}
\includegraphics[clip = true, width=0.9\linewidth]{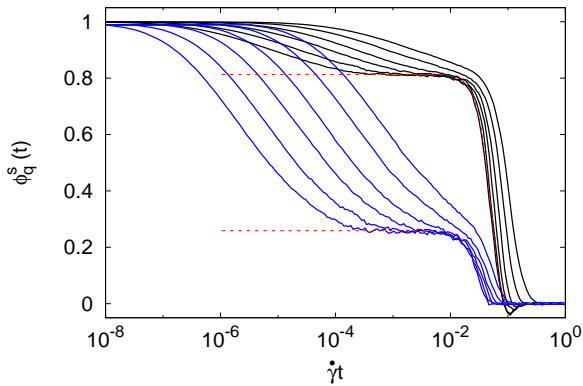}
\caption{Transient incoherent density correlators for $\eta= 0.81$ (glass) and $|{\bm q}|d=6.5$ (upper curves)  and $|{\bm q }|d = 16.22$ (lower curves), for the direction  $q_x=q_y$, rescaled by $\dot \gamma t$. The correlators approach  master functions for Pe$_0\to 0$, which have been fitted by Eq.~\eqref{eq:fit} (dashed lines). \label{fig:phi0.81-shear-w41-strain-diag}}
\end{figure}

As discussed in section \ref{sec:alpha}, MCT-ITT predicts the approach of a master-curve for small shear rates in glassy states (or more precisely in states with Pe$_0\ll1$ and Pe$\gg1$), which depend on time only via accumulated strain $\dot\gamma t$. In Fig.~\ref{fig:phi0.81-shear-w41-strain-diag}, we demonstrate that the simulation curves indeed approach a master function for the system at $\eta= 0.81$, exemplarily for the direction $q_x=q_y$; similar behavior for the other directions can be observed in Fig.~\ref{corr-simu-dir-glass}.

When comparing the properties of the master-curves in detail, we first have to note that in both theory and simulation, only parts of the $\alpha$-process can be well fitted by a compressed exponential, Eq.~\eqref{eq:fit}. On the theoretical side, the direction $q_x=0$ is an exception, since it can be well fitted by a compressed exponential throughout the $\alpha$-process, compare Fig.~\ref{fig:incohgl}, and we will use this direction for the comparison. On the simulation side, the curves are almost isotropic up to $\dot\gamma t=0.1$ and, e.g.,  the $q_x=q_y$ direction can be well described by a compressed exponential except for the very last part becoming negative and then oscillating back. These fits describe all directions up to the point were the shoulders emerge.  

\begin{figure}
\includegraphics[clip = true, width=0.9\linewidth]{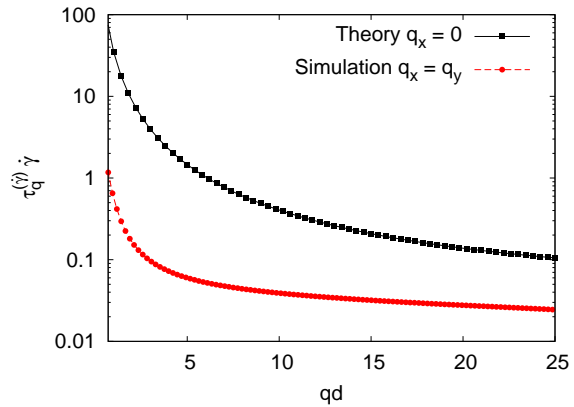}
\caption{The relaxation timescale $\tau^{(\dot\gamma)}_q$ of the master function $(\eta= 0.81)$ as function of wavevector. See main text for a discussion of the differences. \label{fig:tausim}}
\end{figure}

In Fig.~\ref{fig:tausim}, we show the comparison of the relaxation timescale obtained from this fitting procedure. The theoretical curve is identical to the one in Fig.~\ref{fig:tau}. For the simulations, the fit was done with our smallest shear rate (Pe$_0=2 \times 10^{-4}$). We see that, while the overall shape of the time scale as function of $q$ is the same in theory and simulation, the theory overestimates the relaxation time by about a factor of 70 for the smallest $q$. For large $q$, the agreement is much better as there is roughly a factor of $5$ difference. For $q=6.6$, the difference is roughly a factor 20, i.e., more than 4 times larger then the deviation for the initial decay from the plateau (compare Figs.~\ref{corr-simu-dir-glass} and \ref{fig:betafsim}). This additional factor can hence be attributed to the underestimation of the compression in the final decay by theory. 

\begin{figure}
\includegraphics[clip = true, width=0.9\linewidth]{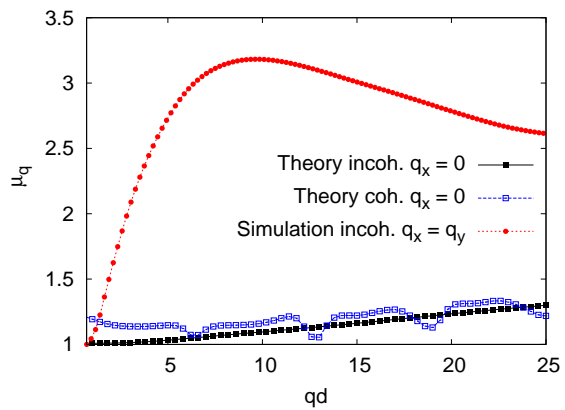}
\caption{\label{fig:betasim}The compressing exponent $\mu_{\bf q}$ of the master function $(\eta= 0.81)$ as function of wavevector.}
\end{figure}

This line of argument carries over to Fig. \ref{fig:betasim}, where we compare the stretching exponent (in our case rather a compressing exponent) for the master functions. As discussed before, the fact that the exponent is larger than unity can be interpreted as a signature of non-stationarity, as it seems to only appear for transient quantities \cite{Krueger10b,Zausch08}. While MCT-ITT correctly captures this nontrivial feature on a qualitative basis, the exponent is much larger in the simulations. We additionally see that the exponent in the simulations has a maximum as function of $q$, which can again be understood as a consequence of the stress overshoot: For large $q$, the functions have relaxed to zero before the overshoot sets in (compare the timescales in Fig.~\ref{fig:tausim}), hence they do not feel the effect of overshoot and are less compressed. As mentioned above, this effect seems not to be captured by our theory, as the exponents increase steadily with $q$. Further evidence is given in Fig.~\ref{fig:tausim} where theory and simulation approach each other for large $q$, where the overshoot effect plays no role.
We also want to emphasize that the direction $q_x =0$ has for most cases the least steep curves, for other directions, we observe exponents as large as nearly 2 in theory. 

\begin{figure}
\includegraphics[clip = true, width=0.9\linewidth]{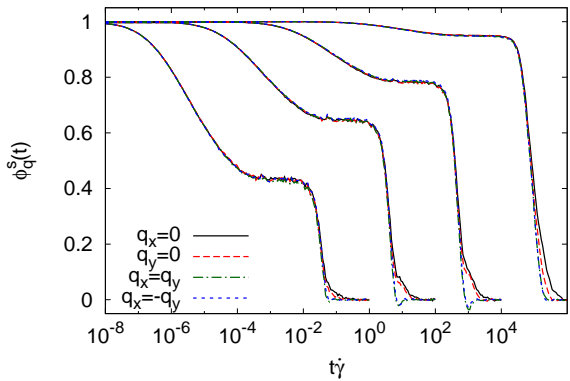}
\caption{Transient incoherent correlators at $\eta=0.81$ (glass) for the directions as labeled, for wavevectors $|{\bm q}|d =2.99$, $|{\bm q}|d =6.63$, $|{\bm q}|d =8.96$ and $|{\bm q}|d = 12.6$ from right to left. All correlators are at the lowest shear rate accessible Pe$_{0}= 2 \times 10^{-4}$, but shifted in time by a factor $\dot \gamma =10^0$, $10^{-2}$, $10^{-4}$ and $10^{-6}$ from left to right for visibility.
} \label{fig:phi_s81-varq-strain}
\end{figure}

Finally, Fig.~\ref{fig:phi_s81-varq-strain} shows the master-curve for $\eta=0.81$ (obtained for Pe$_0 = 2 \times 10^{-4}$) for different wave vectors (compare Fig.~\ref{fig:allangles}), allowing to study the wave vector dependence of the shoulders. For small $q$, the shoulders take up about $40 \%$ of the $\alpha$ process, while for larger $q$, they only emerge during the last $20 \%$ of the final relaxation (estimated from comparing the height of the point, where the correlators start to split, to the plateau height). As in theory (Fig.~\ref{fig:allangles}), we observe that the shoulders are a cooperative effect, happening for all wavevectors roughly at the same time $\dot\gamma t\approx0.07$  (the time corresponding to the slowing down of the system after stress relaxation), and we expect that the effect vanishes for large $q$ (where the correlators are zero when this happens). Comparing with Fig.~\ref{fig:allangles}, we note that MCT-ITT slightly overestimates the  shoulders  for small $q$.

\subsection{Mean Squared Displacements}\label{sec:simMSD}
\begin{figure}
\includegraphics[clip = true, width=0.9\linewidth]{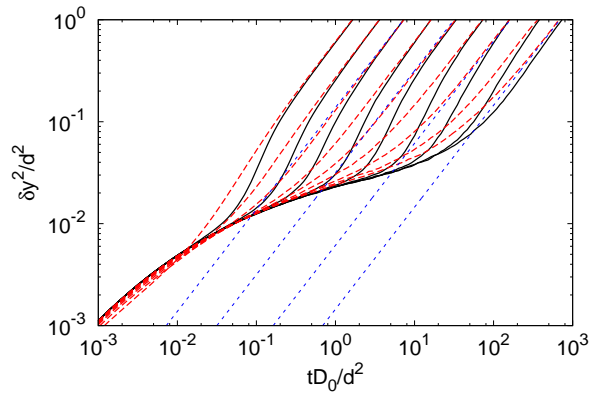}
\caption{Comparison of stationary (dashed)  and transient (solid) MSDs for the gradient direction at $\eta=0.79$. Pe$_0$ numbers are  $2 \times 10^0, 6 \times 10^{-1},  2 \times 10^{-1},  6 \times 10^{-2}, 2 \times 10^{-2}, 6  \times 10^{-3}$, $2 \times 10^{-3},  6 \times 10^{-4} ,2 \times 10^{-4}$ from left to right. Thin dashed lines show fits to the  asymptote $2 D_y^{(\dot \gamma)} t$. \label{fig:msd-y-phi0.79}}
\end{figure}

Let us finally discuss the mean squared displacements.
Given the definitions   $\delta x_{i_j}(t_k,t_w) = x_{i_j}(t_k+t_w) - \dot \gamma  t_k y_{i_j} (t_k+t_w) -   x_{i_j}(t_w) $ and $\delta y_{i_j}(t_k,t_w) =y_{i_j}(t_k+t_w) -  y_{i_j}(t_w) $ for the displacement at time-step $t_k$ for a particle $i$ in $x,y$-direction for the $j$-th simulation run, we define the mean squared displacements in a similar manner as in Eqs.~\eqref{eq:msdy}, \eqref{eq:msdx} and \eqref{eq:xydef}. Again $t_w=0$ and $\dot\gamma t_w=1$ for transient and stationary cases, respectively,  
\begin{align}
 \delta y^2  (t_k, t_w) &= \frac{1}{MN} \sum \limits_{j=1}^M \sum \limits_{i=1}^N \delta y_{i_j}(t_k, t_w)  \delta y_{i_j}(t_k, t_w) ,\\
\delta x^2  (t_k,t_w ) &= \frac{1}{MN} \sum \limits_{j=1}^M  \sum \limits_{i=1}^N \delta x_{i_j}(t_k,t_w )  \delta x_{i_j}(t_k, t_w ) ,\\
 \delta x y  (t_k, t_w) &= \frac{1}{MN} \sum \limits_{j=1}^M  \sum \limits_{i=1}^N \delta x_{i_j}(t_k,t_w)  \delta y_{i_j}(t_k, t_w) .
\end{align}
The difference between transient and stationary curves have been discussed in Ref.~\cite{Krueger10b} and demonstrated in simulations for the liquid case. 
Here we show these curves again for completeness in Fig.~\eqref{fig:msd-y-phi0.79}. We see that stationary and transient correlators approach each other for long times. This supports the argument given in Sec.~\ref{sec:MSDstat} that our result \eqref{eq:ydir} indeed gives the long time diffusivity of the steady state.

\begin{figure}
\includegraphics[clip = true, width=0.9\linewidth]{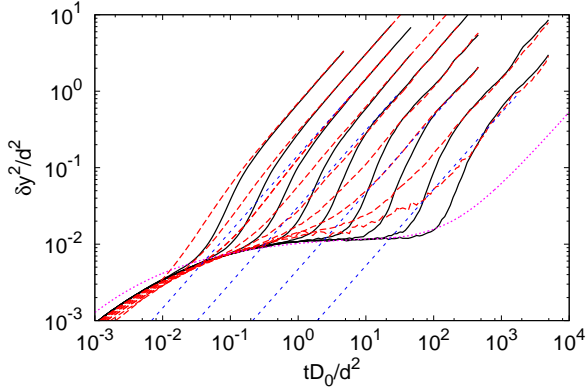}
\caption{Stationary (dashed) and transient (solid) MSDs for the gradient direction at $\eta=0.81$ (glass). Pe$_0$ numbers are  $2 \times 10^0, 6 \times 10^{-1},  2 \times 10^{-1},  6 \times 10^{-2}, 2 \times 10^{-2}, 6  \times 10^{-3}$, $2 \times 10^{-3},  6 \times 10^{-4} ,2 \times 10^{-4}$ from left to right. The dotted (magenta) curve is the theoretical result for Pe$_0=10^{-3}$ and $\varepsilon=10^{-3}$.  Thin dashed lines show fits to the asymptote $2 D_y^{(\dot \gamma)} t$. \label{fig:msd-y-phi0.81}}
\end{figure}

Fig.~\ref{fig:msd-y-phi0.81} shows the same plot for the glassy case which was not presented in Ref.~\cite{Krueger10b}. Here we can test another prediction from the theory; The difference between stationary and transient curves prevails up to arbitrarily small shear rates (as long as Pe$\gg1$ holds). This is a nontrivial statement in agreement with Fig.~\ref{fig:msdstat}. There is as yet a qualitative difference between theory and simulations concerning the transient curves. The simulations show superdiffusive behavior connected to the stress overshoot \cite{Krueger10b,Zausch08} which is underestimated in the theory (Figs.~\ref{fig:msd} and \ref{fig:msdstat}), as discussed before. One possibility for $\delta y^2$ in Eq.~\eqref{eq:ymsdeq} to show superdiffusive behaviour is a negative memory function at long times \cite{Zausch08}. But the memory function in Eq.~\eqref{eq:ymsdeq} numerically turns out to be positive. There is no mathematical reason for this positivity inherent in the structure of our equations, and indeed it seems simple coincidence that $m_y^0(t)$ is positive for all $t$: Changing its structure slightly can lead to negative values for long times and yield superdiffusive motion.

As was the case for the relaxation time scales in Fig.~\ref{fig:tausim}, this underestimation of our theory gives rise to rather large deviations of the long time diffusivities from the simulation values. In Fig.~\ref{fig:msd-y-phi0.81}, we additionally show the theoretical transient curve for Pe$_0=10^{-3}$ demonstrating a scenario equivalent to Fig.~\ref{corr-simu-dir-glass}. While there is in principal no free parameter in our theory, we multiply both axis of the theoretical data by a factor of 1.22. This factor sets the plateau values equal, which are naturally slightly different in the binary mixture compared to our mono-disperse theory. It has no effect on the timescales of the curves which we want to discuss here: We see that the theory curve for Pe$_0=10^{-3}$ leaves the glassy plateau at the same time as the simulation curve for Pe$_0=2\times10^{-4}$, in agreement to what we observed in Fig~\ref{corr-simu-dir-glass}. Again, up to this point, theory and simulation agree up to a factor of less than 5 in Pe$_0$. Going to larger times, the memory function $m_y^0(t)$ in Eq.~\eqref{eq:ymsdeq} misses to become negative and the theory curves are not steep enough. As seen in the plot, the long time diffusivity differs then by roughly a factor $55$ (hence roughly 10 times more then initially). Regarding again the result for the long time diffusivity in Eq.~\eqref{eq:ydir}, one sees that a negative part in $m_y^0(t)$ could possibly render $m_y^0(z=0)$ much smaller giving larger values for the diffusivities.

\begin{figure}
\includegraphics[clip = true, width=0.9\linewidth]{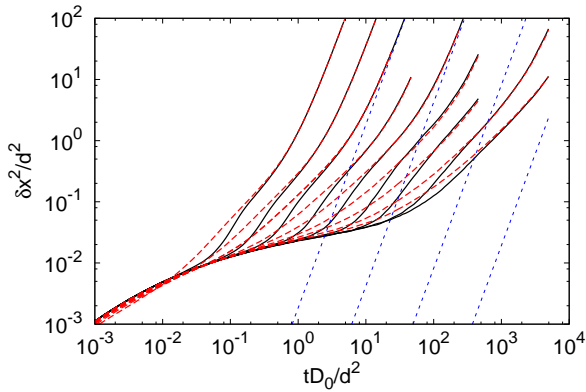}
\caption{Stationary (dashed) and transient (solid)  MSDs for the flow direction at $\eta=0.79$ (liquid). Pe$_0$ numbers are $2 \times 10^0, 6 \times 10^{-1},  2 \times 10^{-1},  6 \times 10^{-2}, 2 \times 10^{-2}, 6  \times 10^{-3}$, $2 \times 10^{-3},  6 \times 10^{-4} ,2 \times 10^{-4}$ from left to right. The asymptotes $2/3 D^{(\dot \gamma)}_y t^3 \dot \gamma^2$, using the same $D_y^{(\dot \gamma)}$ as in Fig.~\ref{fig:msd-y-phi0.79} for the corresponding shear rates, are shown as thin dashed lines.    \label{fig:msd-x-phi0.79}}
\end{figure}

\begin{figure}
\includegraphics[clip = true, width=0.9\linewidth]{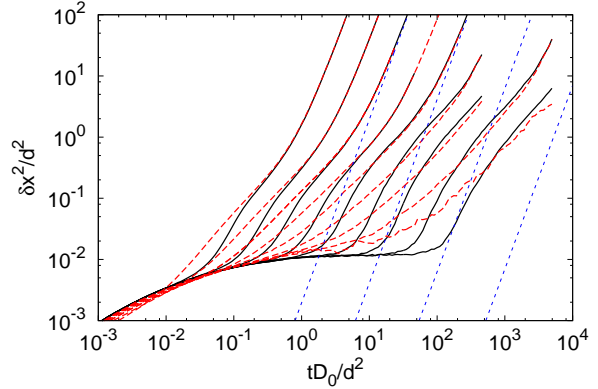}
\caption{Stationary (dashed) and transient (solid)  MSDs for the flow direction at $\eta=0.81$ (glass). Pe$_0$ numbers are  $2 \times 10^0, 6 \times 10^{-1},  2 \times 10^{-1},  6 \times 10^{-2}, 2 \times 10^{-2}, 6  \times 10^{-3}$, $2 \times 10^{-3},  6 \times 10^{-4} ,2 \times 10^{-4}$ from left to right. The asymptotes $2/3 D^{(\dot \gamma)}_y t^3 \dot \gamma^2$, using the same $D_y^{(\dot \gamma)}$ as in Fig.~\ref{fig:msd-y-phi0.81} for the corresponding shear rates, are shown as thin dashed lines.  \label{fig:msd-x-phi0.81}}
\end{figure}

There are three more things to test in our simulations regarding the flow direction presented in Figs.~\ref{fig:msd-x-phi0.79} (liquid) and \ref{fig:msd-x-phi0.81} (glass). First, we note that the simulations indeed show the glass Taylor dispersion, the MSDs grow proportional to $t^3$ for long times. Second, transient and stationary curves also merge for the flow direction at long times, i.e., the expression \eqref{eq:Taylor1} holds for both transient and stationary curves as argued in Sec.~\ref{sec:MSDstat}. Third, our simulations indeed confirm the nontrivial statement of Eq.~\eqref{eq:Taylorgl} for both liquid and glass: The $t^3$ term is connected to the diffusivity for the $y$ direction as in the low density case, Eq.~\eqref{eq:elrick}. In 3D systems, where diffusivities are slightly anisotropic for the two directions perpendicular to the flow \cite{Zausch08}, we predict that the $t^3$-term is connected to the gradient direction rather than the neutral direction, as expressed by Eq.~\eqref{eq:Taylorgl}.

\begin{figure}
\includegraphics[clip = true, width=0.9\linewidth]{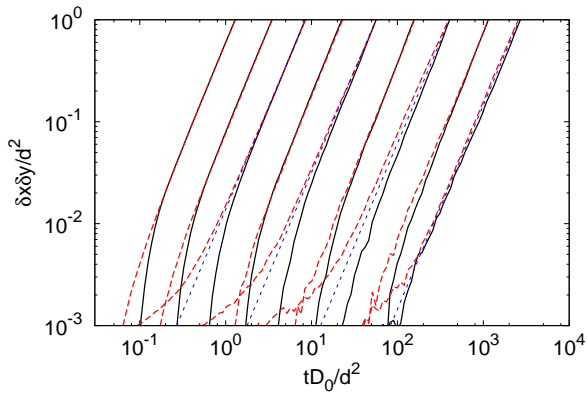}
\caption{Stationary (dashed) and transient (solid)  cross correlators at $\eta=0.79$ (liquid). Pe$_0$ numbers are $2 \times 10^0, 6 \times 10^{-1},  2 \times 10^{-1},  6 \times 10^{-2}, 2 \times 10^{-2}, 6  \times 10^{-3}$, $2 \times 10^{-3},  6 \times 10^{-4} ,2 \times 10^{-4}$ from left to right.  
The asymptotes $ D^{(\dot \gamma)}_y \dot  \gamma t^2$, using the same values of $ D^{(\dot \gamma)}_y$ as in Fig.~\ref{fig:msd-y-phi0.79} for the respective shear rates, are shown as thin dashed lines.
\label{fig:msd-xy-phi0.79}}
\end{figure}

\begin{figure}
\includegraphics[clip = true, width=0.9\linewidth]{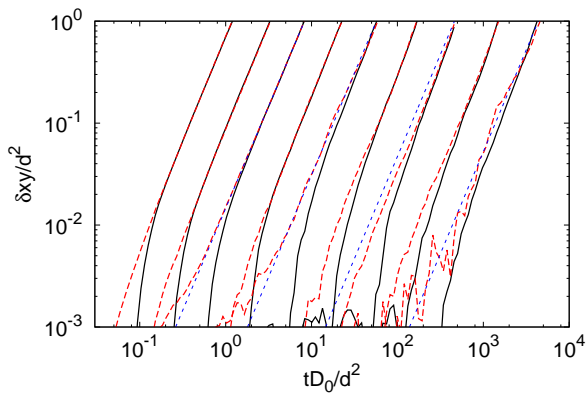}
\caption{Stationary (dashed) and transient (solid)  cross correlators at $\eta=0.81$ (glass). Pe$_0$ numbers are  $2 \times 10^0, 6 \times 10^{-1},  2 \times 10^{-1},  6 \times 10^{-2}, 2 \times 10^{-2}, 6  \times 10^{-3}$, $2 \times 10^{-3},  6 \times 10^{-4} ,2 \times 10^{-4}$ from left to right.  The asymptotes $ D^{(\dot \gamma)}_y \dot  \gamma t^2$, using the same values of $ D^{(\dot \gamma)}_y$ as in Fig.~\ref{fig:msd-y-phi0.81} for the respective shear rates, are shown as thin dashed lines. \label{fig:msd-xy-phi0.81}}
\end{figure}

Inspecting the cross correlator $\delta x y (t,t_w)$, as shown in Figs.~\ref{fig:msd-xy-phi0.79} and \ref{fig:msd-xy-phi0.81} for liquid and glass, respectively, we can confirm further predictions of the theory. Again for long times transient and stationary functions coincide as expected from section~\ref{sec:MSDstat}. Furthermore the connection between  shear and gradient directions can be seen, as the long time asymptote (shown as blue lines) $D^{(\dot \gamma)}_y \dot  \gamma t^2$ uses the same $D^{(\dot \gamma)}_y$ as in Figs.~\ref{fig:msd-y-phi0.79} and \ref{fig:msd-y-phi0.81} for the corresponding shear rates. This confirms the theoretical prediction expressed in Eq.~\eqref{eq:xyres}.

\begin{figure}
\includegraphics[clip = true, width=0.9\linewidth]{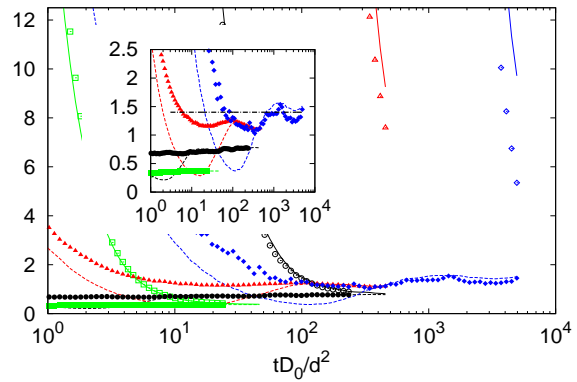}
\caption{The quantities $3\delta x^2 (t,t_w) / (2\dot \gamma^3 t^3)$  and $\delta y^2(t,t_w) / (2 \dot \gamma t)$, testing their approach of $\beta_y$ in Eqs.~\eqref{eq:scalingx} and \eqref{eq:alphadef} for $t\to\infty$. Symbols denote the stationary, lines the transient curves. Open symbols and solid lines show the shear-, filled symbols and  dashed lines the gradient direction. The Pe$_0$ numbers are $2 \times 10^{-1}$ (squares), $2 \times 10^{-2}$ (circles), $2 \times 10^{-3}$ (triangles) and $2 \times 10^{-4}$ (diamonds). For the two largest shear rates, collapse of shear and gradient directions for long times is visible. The inset shows the curves for the gradient direction magnified where convergence of $\beta_y(\dot\gamma)$ to the value of the master-curve for Pe$_0\to0$ (indicated by a horizontal black line) is observed.
    }\label{fig:msd-collapse}
\end{figure}

As a further and more sensitive test of the scaling property in Eqs.~\eqref{eq:scalingx} and \eqref{eq:alphadef}, the quantities $3\delta x^2 (t,t_w) / (2\dot \gamma^3 t^3)$  and $\delta y^2(t,t_w) / (2t \dot \gamma)$ are shown in Fig.~\ref{fig:msd-collapse}, for the transient and stationary curves. For long times the so defined mean-squared displacements for shear- and gradient direction collapse on the constant $\beta_y=D^{(\dot \gamma)}_y / \dot \gamma$ for the two largest shear rates as expected, while the two lowest shear rates already show the right trend, presumably reaching their asymptote outside the window accessible in the simulation. The inset of Fig.~\ref{fig:msd-collapse} magnifies the gradient direction for both transient and stationary curves, where the superdiffusive regime of the transient mean-squared displacement expresses itself by a dip before reaching the long time asymptote. For decreasing shear rate, the curves approach the scaling constant $\beta_y(\dot\gamma\to0)\approx 1.4$, indicated by the horizontal black line. We emphasize again that this number uniquely describes all possible MSDs (in 2D) for long times.

\begin{figure}
\includegraphics[clip = true, width=0.9\linewidth]{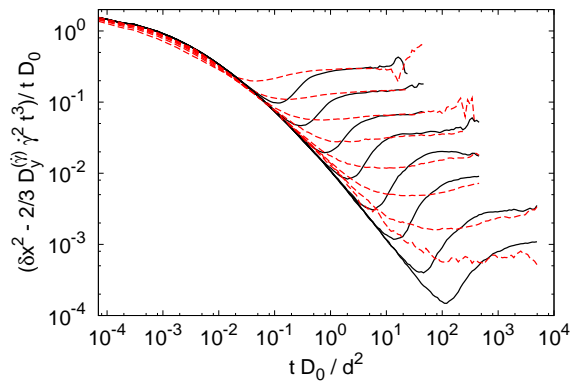}
\caption{Investigation of the next to leading term in the MSD for the shear direction in simulations at $\eta=81$ (glass). Pe$_0$ numbers are  $6 \times 10^{-1},  2 \times 10^{-1},  6 \times 10^{-2}, 2 \times 10^{-2}, 6  \times 10^{-3}$, $2 \times 10^{-3},  6 \times 10^{-4} ,2 \times 10^{-4}$ from left to right. Shown are transient (solid) and stationary (dashed) MSDs after subtraction of the leading term $\propto t^3$ and subsequent division by $t$. }\label{fig:nextorder}
\end{figure}

In theory, the MSD for the shear direction contains a term of order $\mathcal{O}(t^2)$ (compare Eq.~\eqref{eq:bt}).  Fig.~\ref{fig:nextorder} shows this MSD in simulations, after subtraction of the $t^3$-term and division by $t$. A term of order $\mathcal{O}(t^2)$, which  would manifest itself in a linear increase of the curves at long times, however, cannot be resolved. 

\begin{figure}
\includegraphics[clip = true, width=0.9\linewidth]{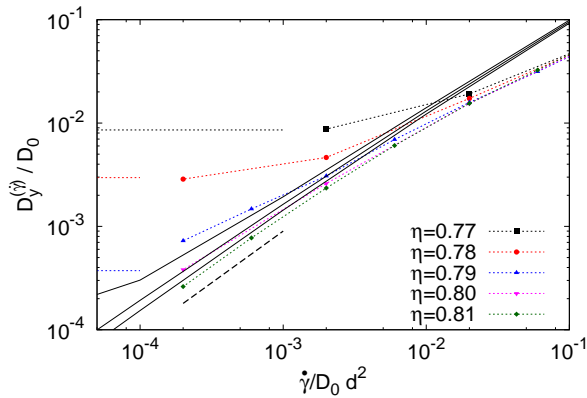}
\caption{Long time diffusion coefficient as a function of Pe$_0$ for different densities. Solid lines show the theoretical data for $\varepsilon=-10^{-3},0,10^{-3}$ from top to bottom. Theoretical data has been shifted by a factor of 55 along the $x$-axis. A dashed bar shows the slope of 1. Dashed horizontal lines show the Pe$_0\to 0$ asymptote from the quiescent system. 
\label{fig:diffusion}}
\end{figure}

In Fig.~\ref{fig:diffusion}, we finally show the long time diffusion coefficients for the $y$-direction (as defined in Eq.~\eqref{eq:alphadef}), as a function of shear rate. For large shear rates, the diffusivities for the different densities are very close together, a behavior which is known also from the macroscopic shear viscosities \cite{Fuchs03}. As shear rate gets smaller, the diffusivities for the liquid densities finally approach a constant value given by the diffusivity of the unsheared suspension. These decrease with density \cite{Bayer07}. On the glassy side, we observe the approach of the scaling regime, where the diffusivities are linear in shear rate. Simulation and theory agree with respect to all these findings. Quantitatively, there is a factor of roughly 55 between theory and simulation, see the discussion of Fig.~\ref{fig:msd-y-phi0.81}.

\section{Summary}
We discussed some of the characteristic features of tagged particle dynamics for glasses under shear. The transient tagged particle density correlator shows strong imprints from the shear stress after switch on. Directly after the stress-overshoot, the correlation functions decay quickly and are superexponential. Nevertheless, after the stress has relaxed to its final value, they drastically slow down, leading to the appearance of a direction-dependent final shoulder. Despite the strong anisotropy of the applied flow field, the correlation functions show rather small effects of anisotropy. The mean square displacement of the tagged particle shows an effect known at low density  as Taylor dispersion, which at the glass transition appears in modified form to obey the scaling with shear rate. The coupling of the MSDs for shear and gradient directions is identical to the low density case.

The extension of mode coupling theory to sheared systems (MCT-ITT) allows to study the properties of the tagged particle correlator and the MSD. It captures many nontrivial effects (e.g. anisotropy of correlation functions, superexponential behavior, emergence of shoulders, scaling behavior in the glass for both stationary and transient functions, Taylor dispersion), and gives quantitative predictions without adjustable parameters, where the resulting timescales are captured correctly within roughly one order of magnitude. We attribute the deviations in timescales to an underestimation of the stress-overshoot scenario in theory, as both correlators and MSDs do not speed up as strongly as the simulation curves after the stress maximum is passed.
\section{Acknowledgments}
We thank J.~F.~Brady, J.~M. Brader and T.~Voigtmann for discussions. 
M. K. was supported by the Deutsche Forschungsgemeinschaft via the SFB-TR6 and grant KR 3844/1-1, and FW acknowledges partial support by the German Excellence
Initiative.

\begin{appendix}

\section{$\alpha$-scaling equation}\label{app:alpha}
To derive the equation for $\Phi^{s+}_{\bf q}(\tilde t)$, we start with the convolution integral in Eq.~\eqref{eq:incofin},
\begin{equation}
\int_0^t m_{\bf q}^{s}(t,t')\,\partial_{t'} \Phi_{\bf q}^{s}(t').\label{eq:alpha1}
\end{equation}
In Eq.~\eqref{eq:mfin2}, we saw explicitly how the memory function depends on the two different times, namely by 
\begin{equation}
m_{\bf q}^{s}(t,t')=\bar m_{\vct{q}(\dot\gamma t')}^{s}(t-t').
\end{equation}
It depends only on accumulated strain $\dot\gamma t'$ rather than on $ t'$. It does so via the advected wavevector. That means that the dependence on $t'$ is already  $\alpha$-scaling-like. Using this, we can rewrite the integral in Eq.~\eqref{eq:alpha1} to
\begin{align}
&-\bar m_{\bf q}^{s}(t)+\frac{d}{dt}\int_0^t dt' \bar m_{\vct{q}(\dot\gamma t')}^{s}(t-t')\Phi_{\bf q}^{s}(t')\nonumber\\
&-\int_0^t dt' \frac{\partial \vct{q}(\dot\gamma t')}{\partial t'}\cdot\left(\frac{\partial}{\partial \vct{q}(\dot\gamma t')}\bar m_{\vct{q}(\dot\gamma t')}^{s}(t-t')\right) \Phi_{\bf q}^{s}(t').
\end{align}
With this, the $\alpha$ scaling equation, Eq.~\eqref{eq:alphasc}, follows.
\section{Long time solution of the MSDs}\label{app:taylor}

It is useful to rewrite the memory function $m_{\bf q}^{s}(t,t')$ into a product of the part which depends only on the difference $t-t'$ and the part which depends explicitly on $t$ and $t'$,
\begin{equation}
m^{s}_{\bf q}(t,t')=\sum_{{\bf k}}\frac{\vct{k}(t-t')\cdot\vct{q}(t)}{q^2(t)}\frac{\vct{k}\cdot\vct{q}(t')}{q^2(t')} \, F(\vct{k},\vct{q}(t'),t-t'),\label{eq:mfMSD}
\end{equation}
where the function \begin{align}&F(\vct{k},\vct{q}(t'),t-t')=\notag\\&\frac{1}{N}n^2 c_{k(t-t')}^{s}c_{k}^{s} S_{k}\,\Phi_{\vct{k}-\vct{q(t')}}^{s}(t-t')\Phi_{\vct{k}}(t-t')\label{eq:mfMSD2}\end{align} depends still explicitly on $t'$ via the wavevector $\vct{q}(t')$. However, this dependence will vanish in the low $q$ limit as used for the calculation of the MSDs.
\subsection{Flow direction}
In order to find the long time solution of Eq.~\eqref{eq:xricht} we write
\begin{equation}
\lim_{t\to\infty}\delta x^2(t)=a t^3+b t^2+c t+\dots.\label{eq:ansatz}
\end{equation}
The form \eqref{eq:ansatz} can be justified knowing that the function $F$ in Eq.~\eqref{eq:mxricht} decays to zero as $t-t'\to\infty$. A term $t^4$ (or higher powers in $t$, or fractional powers) does not exist since the initial decay rate \eqref{eq:init} does not contain such a term. The long time behavior is hence governed by the initial decay rate, a fact which is interesting because the long time behavior of the correlator $\Phi_{\bf q}^s(t)$ is independent of the initial decay rate as $\dot\gamma\to 0$ \cite{Fuchs03}. This is because the limits of $t\to\infty$ and $q\to 0$ do not commute. \\
We first determine the coefficient $a$. For this, the leading long time (large $t'$ and $t''$) terms in the integral in Eq.~\eqref{eq:xricht} are needed. They are independent of the coefficient $b$. The equation for $b$, on the other hand, will contain the coefficient $a$ and can hence only be solved afterwards. The leading term of the first bracket in $m_x^0(t,t')$ is $-\dot\gamma tk_y(t-t')$, the leading term of the fraction is given by $-k_y/(\dot\gamma t')$. With this, we get
\begin{equation}
3 at^2+\int_0^{t}dt'\dot\gamma t m_y^0(t-t')   \frac{\partial_{t'} at'^3}{\dot\gamma t'}=2\dot\gamma^2t^2.\label{eq:longtime}
\end{equation}
We note that $m_y^0(t)$ appears. This equation can be treated with the following formula for Laplace transforms \cite{Doetsch},
\begin{eqnarray}
\mathcal{L}\left\{t f(t)\right\}(z)&=&-\frac{\partial}{\partial z}\mathcal{L}\left\{f(t)\right\}(z).
\end{eqnarray}
Using it, we find that the integral in Eq.~\eqref{eq:longtime} contain also one term of order $t^2$ (because $\partial_z m_y^0(z)$ is finite as $z\to0$), which does not contribute to $a$. We find
\begin{equation}
a=\frac{2\dot\gamma^2}{3+3m_y^0(z=0)}.\label{eq:a}
\end{equation}
We must also consider the next order leading term as it will be needed in order to calculate the $xy$ cross correlation. For the equation for the coefficient $b$, all long time terms proportional to $t^2$ have to be collected, (note that some of the possible contributions vanish in the sum over $\bf k$ due to symmetries \footnote{Although the memory function including the correlators is not isotropic under shear, it is still symmetric with respect to the origin, $\vct{k}=0$, since the system is symmetric with respect to the origin.}),
\begin{equation}
2b+2bm_y^0(z=0)-3a\partial_z m_y^0(z=0)-3a\frac{m_{xy}^0(z=0)}{\dot\gamma}=0.\label{eq:b}
\end{equation}
The promised dependence of $b$  on $a$ appears. Also, the off-diagonal memory function enters,
\begin{equation}
m_{xy}^0(t)=\sum_{{\bf k}}k_y(t)k_x\, F(\vct{k},\vct{0},t).
\end{equation}
We find for $b$, 
\begin{equation}
b=\frac{3a}{2}\frac{\left[\partial_z m_y^0(z=0)\right]+\frac{m_{xy}^0(z=0)}{\dot\gamma}}{1+m_y^0(z=0)}.
\end{equation}
\subsection{Cross correlation}\label{app:cross}
In order to find the long time solution of Eq.~\eqref{eq:xy2}, we have to find the long time behavior of
\begin{equation}
\lim_{q\to0}\frac{1-\Phi_{q(\vct{e}_x+\vct{e}_y)}^{s}(t)}{q^2}\equiv\tilde\delta xy(t)/2.
\end{equation}
Its equation of motion is given by, 
\begin{eqnarray}
&&\partial_t \tilde\delta xy+\int_0^{t}dt'\sum_{\vct{k}}\left[k_x+(1-\dot\gamma t)k_y(t-t')\right] \notag\\&&\frac{k_x+(1-\dot\gamma t') k_y}{1+(1-\dot\gamma t')^2}F(\vct{k},\vct{0},t-t') \partial_{t'} \tilde\delta xy(t')\nonumber\\&&=2\frac{\Gamma^s_{q\vct{e}_x+{q\vct{e}_y}}(t)}{q^2}.\label{eq:deltaxy}
\end{eqnarray}
Again, we can only solve this equation for the long time contributions after making the following ansatz,
\begin{equation}
\lim_{t\to\infty}\tilde\delta xy(t)=a't^3+b't^2+c't+\dots.\label{eq:series}
\end{equation}
We note that the leading long time term in Eq.~\eqref{eq:deltaxy} is equal to the long time term of $\delta x^2$, i.e.,
\begin{equation}
a'=a.
\end{equation}
Additionally to the terms in Eq.~\eqref{eq:b}, the equation for $b'$ contains one contribution from the initial decay rate. The additional terms in Eq.~\eqref{eq:deltaxy} that come from the memory function exactly cancel each other in this order. We find for $b'$,
\begin{equation}
b'=b-\frac{2\dot\gamma}{1+m_y^0(z=0)}.\label{eq:bp}
\end{equation}
It is important that $a'=a$, leading to the cancellation of the $t^3$-terms in Eq.~\eqref{eq:xy2}. Putting the result of Eq.~\eqref{eq:series} into Eq.~\eqref{eq:xy2} leads to Eq.~\eqref{eq:xyres}.
\end{appendix}

\end{document}